\documentclass[a4paper,11pt]{article}
\pdfoutput=1 
\usepackage{jheppub} 
\usepackage[T1]{fontenc} 
\usepackage{amsmath}
\usepackage{amssymb}
\usepackage{amsfonts}
\usepackage{graphicx}
\usepackage{enumitem}
\usepackage{bm}
\usepackage{rotating}
\usepackage{hyperref}
\usepackage{pbox}
\usepackage[dvipsnames]{xcolor}
\usepackage{array}
\usepackage{physics}
\usepackage{bm}
\usepackage{dsfont}
\usepackage{mathtools}
\usepackage{leftidx}
\usepackage{lmodern}
\usepackage{braket}
\usepackage{tensor}
\usepackage{mathrsfs}
\usepackage{float}
\usepackage[title]{appendix}
\usepackage[normalem]{ulem}

\usepackage{tikz}
\usetikzlibrary{arrows.meta,decorations.pathmorphing,math}

\newcommand{\tcr}[1]{\leavevmode{\color{red}{#1}}}

\newcommand{\pstar}{\eta}

\numberwithin{equation}{section}

\newcommand{\be}{\begin{equation}}
\newcommand{\ee}{\end{equation}}
\newcommand{\ba}{\begin{eqnarray}}
\newcommand{\ea}{\end{eqnarray}}

\newcommand{\beq}{\begin{equation}}
\newcommand{\eeq}{\end{equation}}
\newcommand{\beqa}{\begin{eqnarray}}
\newcommand{\eeqa}{\end{eqnarray}}

\newcommand{\Deff}{\Delta_{\text{eff}}}

\title{Love symmetry in higher-dimensional rotating black hole spacetimes}

\date{today}

\author[a]{Finnian Gray,}
\author[b]{Cynthia Keeler,}
\author[c]{David Kubiz\v n\'ak,}
\author[d]{Victoria Martin}

\affiliation[a]{University of Vienna, Faculty of Physics, Boltzmanngasse 5, A 1090 Vienna, Austria}
\affiliation[b]{Department of Physics, Arizona State University, Tempe, AZ 85281, USA}
\affiliation[c]{Institute of Theoretical Physics, Faculty of Mathematics and Physics,
Charles University, Prague, V Hole{\v s}ovi{\v c}k{\' a}ch 2, 180 00 Prague 8, Czech Republic}
\affiliation[d]{Department of Physics, University of North Florida, Jacksonville, FL 32224, USA}

\emailAdd{finnian.gray@univie.ac.at}
\emailAdd{keelerc@asu.edu}
\emailAdd{david.kubiznak@matfyz.cuni.cz}
\emailAdd{victoria.martin@unf.edu}

\abstract{
We develop a method for constructing a 1-parameter family of
globally-defined Love symmetry generators in rotating black hole spacetimes of  general dimension. 
The key ingredient is to focus on the vicinity of the (physical) outer horizon, matching only the radial derivative and the outer horizon pole pieces of the Klein--Gordon operator in the black hole spacetime to the $SL(2,\mathbb{R})$ Casimir operator.  
After revisiting the 4D Kerr and 5D Myers--Perry cases, the procedure is illustrated on generalized Lense--Thirring spacetimes which describe a wide variety of slowly rotating black hole metrics in any number of dimensions. 
Such spacetimes are known to admit an extended tower of Killing tensor and Killing vector symmetries and, as demonstrated in this paper, allow for separability of the massive scalar wave equation in  Myers--Perry-like coordinates. 
Interestingly, separability also occurs  in the horizon-penetrating Painlev{\'e}--Gullstrand coordinates associated with the freely infalling observer who registers flat space around her all the way to singularity.

}

\begin{document} 

\maketitle
\flushbottom

\section{Introduction}

The Kerr/CFT correspondence \cite{Guica:2008mu} is an enticing conjecture positing that the near-horizon geometry of an extremal Kerr black hole admits a 2D CFT description. 
The conjecture is bolstered by the fact that the Cardy formula for the proposed dual CFT exactly reproduces the Bekenstein--Hawking entropy of the Kerr black hole. 
This conjectured duality was quickly extended to an ``extremal black hole/CFT correspondence'' that exists for black holes in more general theories and in higher spacetime dimensions \cite{Lu:2008jk,Hartman:2008pb,Chow:2008dp,Lu:2009gj}. 

Strong evidence was put forth in \cite{Castro:2010fd} that the near-horizon dynamics of a probe scalar field propagating on a \textit{non-extremal} Kerr black hole background also enjoys a dual 2D CFT interpretation. 
This phenomenon was dubbed {\em ``hidden'' conformal symmetry}, since the symmetry structure is in {the dynamics of} the wave equation rather than the background geometry. 
The notion of hidden conformal symmetry has since been diagnosed in a large class of black holes in 4 and 5 spacetime dimensions \cite{Krishnan:2010pv,Bertini:2011ga,Cvetic:2011hp,Sakti:2020jpo,Chanson:2022wls} and for spherically symmetric, static black holes in general dimension in \cite{Ortin:2012mt}. 
In this work, we demonstrate how to construct hidden conformal symmetry generators for rotating, non-spherically symmetric black holes in general spacetime dimensions for the first time. 

One unsavory fact regarding hidden conformal symmetry generators in the works cited above is that they are generally not {\em globally well-defined} {(except that of a Schwarzschild black hole \cite{Bertini:2011ga})}.
However, in recent years, a globally-defined version of the hidden conformal symmetry generators has been proposed for Kerr \cite{Charalambous:2021kcz} and the {5d} Myers--Perry black hole \cite{Charalambous:2023jgq}. 
This symmetry has been dubbed {\em Love symmetry}, as it offers an explanation for the apparent selection rule governing the vanishing of static tidal Love numbers in four (but not generally in higher) spacetime dimensions. 
The tidal Love numbers for five-dimensional black hole spacetimes have been studied in \cite{Charalambous:2023jgq, Rodriguez:2023xjd}. 
We will always refer to globally-defined hidden conformal symmetry as Love symmetry.

In this work we outline a systematic procedure for obtaining these Love generators for stationary black hole spacetimes in all dimensions.
As a warm up, we will test this procedure in the context of the  four-dimensional Kerr black hole and the five-dimensional Myers--Perry spacetime. 
We will then extend our work to the  recently introduced Lense--Thirring spacetimes in~\cite{Baines:2021qaw} and generalized to multiply-spinning class in all dimensions~\cite{Gray:2021toe,Gray:2021roq}.

The generalized Lense--Thirring spacetimes are of particular interest in this context for a number of reasons. 
First, they provide a generic template for slowly rotating black holes in a wide variety of different theories in and beyond Einstein gravity~\cite{Gray:2021roq}.
Second, as spacetimes in their own right they possess a large tower of Killing tensor and Killing vector symmetries, giving rise to the first physically motivated example of black hole spacetimes that have more hidden than explicit symmetries \cite{Gray:2021toe}. 
Third, as demonstrated in this paper, such symmetries lead to a separable massive scalar equation, both in the Myers--Perry-like and the Panlev{\'e}--Gullstrand (PG) infalling coordinates.   
Fourth, by focusing on the slowly rotating Einstein gravity Lense--Thirring spacetime we can construct Love generators which can interpolate between the known higher-dimensional Schwarschild Love generators~\cite{Bertini:2011ga} and the, as yet undetermined, case of arbitrary dimensional Myers--Perry black holes.

To obtain globally-defined Love generators in a higher-dimensional setting, we first restrict ourselves to a particular class of spacetimes. 
We study a massless scalar field equation in a stationary black hole spacetime with a separable scalar field equation%
\footnote{For hidden conformal symmetry in non-separable black ring solutions, see \cite{Chanson:2022wls}.}
\be \label{scalar}
\Box \Psi=\frac{1}{\sqrt{|g|}}\partial_a\Bigl(\sqrt{|g|} g^{ab}\partial_b\Psi\Bigr)=0\,,
\ee 
where $g_{ab}$ is the corresponding spacetime metric and $g=\det (g_{ab})$ stands for its determinant.
We assume that such separability occurs in a specific (Boyer--Lindquist-type) coordinate system with a timelike coordinate $t$, a radial coordinate $r$, and a general number of azimuthal ($2\pi$ periodic) Killing directions $\phi_i$.
We further assume that the (outer and non-extremal) black hole horizon is located at the largest positive real root $r_+$ of some ($r$-dependent) metric function $\Delta(r)$, and that it corresponds to a Killing horizon, generated by the following Killing vector: 
\be \label{xi}
\xi=\partial_t+\partial_{\pstar}\,,\quad \partial_{\pstar}\equiv\sum_{i=1}^m \Omega^i_+ \partial_{\phi_i}\,,
\ee
where $\Omega^i_+$ are the horizon angular velocities.%
\footnote{ The Killing vector $\xi$ plays a central role in black hole thermodynamics; in particular, as was famously noted in \cite{Wald:1993nt}, the horizon entropy is equal to a Noether charge associated with such a Killing field. Naturally, as we shall see, it also plays a key role in the
construction of Love generators.}
While these assumptions may appear restrictive, they hold for a large class of black holes, beyond those that we directly address in this work.

The diagnostic for hidden conformal symmetry (originally written down in \cite{Castro:2010fd}) is that a portion of the radial scalar Laplacian can be identified with an $SL(2,\mathbb{R})$ Casimir $\mathcal{H}^2$.
This symmetry is promoted to Love symmetry (originally written down in \cite{Charalambous:2021kcz}) when the portion of the radial Laplacian to be identified with an $SL(2,\mathbb{R})$ Casimir is constructed from symmetry generators that are globally well-defined. 
The purpose of this work is to present a systematic means of carrying out this matching that is readily extendable to higher dimensions, encompassing a large class of black holes. 

The main addition we propose for the generalized Lense--Thirring black holes is to match the radial Laplacian to $\mathcal{H}^2$ to leading order in $r_+$. 
That is, we carry out an expansion $\Delta=\Deff+\mathcal{O}\bigl([r-r_+]^3\bigr)$, where now $\Deff$ takes the role of $\Delta(r)$.
In addition to this \emph{near horizon limit}, we also always take, as usual, a \emph{near-region limit}, setting 
\be \label{eq: near-region}
\omega r \ll 1\quad \mbox{and}\quad M\omega^{d-3} \ll 1\,.
\ee 
Here, $M$ stands for the mass of the hole, $\omega$ is the frequency of the scalar wave, and $d$ stands for the number of spacetime dimensions.%
\footnote{Compare this to the original near-region limit proposed for 4D Kerr in  \cite{Castro:2010fd}: $\omega r\ll 1$ and $M\omega \ll 1$.} %
We will show that our systematic procedure reproduces (and extends) the results of \cite{Charalambous:2021kcz,Charalambous:2023jgq}. Furthermore, we provide new results for Love symmetry generators for Einstein and Einstein--Maxwell Lense--Thirring spacetimes of general dimension. 
Lastly, we show that the striking similarity in form of Love generators for Kerr and 5D Myers--Perry persists in higher dimensions, hinting at a universal quality of Love symmetry.

This paper is organized as follows. 
In Section \ref{sec: Conf Symmetry}, we demonstrate how to build globally-defined Love generators for generic black hole spacetimes, before applying our methods to the previously studied cases of Kerr and five-dimensional Myers--Perry.   
In Section \ref{sec: LT}, we turn to generalized Lense--Thirring spacetimes. 
{Since these spacetimes are less known we summarize in detail here the outline}. 
We begin by reviewing some of the relevant spacetime properties in Section \ref{subsec: properties}. 
We then demonstrate for the first time that these metrics admit a separable Klein--Gordon equation in Section \ref{subsec: LT Sep}, {their asymptotic charges are discussed in Section \ref{sec: Charges}}, and we determine the general dimension Love symmetry generators in Section \ref{subsec: Love LT}, paying special attention to Einstein and Einstein-Maxwell theories. 
{Finally, in Section~\ref{sec: infalling} we show that the Lense--Thirring spacetimes are separable in infalling Painlev\'e--Gullstrand coordinates}. 
In Section \ref{sec: diss} we discuss our results and future directions. Lastly, we include an Appendix \ref{Appendix: LT} in which we review further details of the Lense--Thirring spacetime, 
{recapitulating the emergence of hidden symmetries from a Killing tower and identifying the underlying separability structure.}

\section{Systematic method for constructing Love generators} \label{sec: Conf Symmetry}

In this section, we lay down the foundation of our formalism for constructing Love symmetry generators. In Section \ref{subsec:hiddenreview}, we introduce the generators inspired by a generalized set of conformal coordinates, as well as their $SL(2,\mathbb{R})$ Casimir $\mathcal{H}^2$.  
In Section \ref{sec: Radial derivs}, we match the radial derivatives of this Casimir to those in the separated, radial Klein--Gordon equation, in an expansion around $r_+$. 
In Section \ref{sec:LoveKerr}, we review hidden conformal symmetry and Love symmetry in the Kerr black hole, and show that our procedure matches known results in the literature \cite{Charalambous:2021kcz}. Likewise, in Section \ref{sec:MP}, we review Love symmetry in the 5D Myers--Perry black hole, and show that our procedure also renders the Love generators in the literature \cite{Charalambous:2023jgq} through a more systematic means. 
We also comment on the striking similarity between the Love generators of Kerr and 5D Myers--Perry black holes. 
This similarity remains in the higher-dimensional Lense--Thirring spacetimes, which we will demonstrate in Section \ref{sec: LT}.

\subsection{Hidden conformal symmetry and conformal coordinates}\label{subsec:hiddenreview}

We now introduce our choice of $SL(2,\mathbb{R})$ generators, followed by a set of conformal coordinates. 
These coordinates will help us describe the hidden conformal symmetry in our spacetimes of interest, as well as propose a set of globally-defined Love symmetry generators. 
This portion of our analysis and presentation takes significant inspiration from the original non-extremal hidden conformal symmetry work of Castro et. al. \cite{Castro:2010fd}.

We begin by considering the $SL(2,\mathbb{R})$ Casimir
\begin{equation}\label{casimir I}
	\begin{split}
		\mathcal{H}^2&=-H_0^2+\frac{1}{2}\left(H_1H_{-1}+H_{-1}H_1\right)\,,\\
	\end{split}
\end{equation}
whose generators satisfy the following algebra:
\begin{equation}\label{algebra}
	\left[H_0,H_{\pm 1}\right]=\mp iH_{\pm1}\,, \qquad \left[H_1,H_{-1}\right]=2iH_0\,.
\end{equation}
We would like to write these generators in a familiar way, which will help streamline our analysis later. 
To this end, we consider the (local) Poincar\'e patch metric of AdS$_3$ 
\begin{equation}\label{poincare}
	ds^2=\frac{L^2}{y^2}\left(dw^+dw^-+dy^2\right).
\end{equation}
The isometry group of AdS$_3$ is famously $SL(2,\mathbb{R})\times SL(2,\mathbb{R})$. 
The Killing vectors of the metric \eqref{poincare} are (denoting $\partial_\pm=\partial_{w^\pm})$
\begin{equation}\label{generators}
	\begin{split}
		H_1&=i\partial_+\,, \qquad H_0=i\bigl(w^+\partial_++\frac{1}{2}y\partial_y\bigr)\,, \qquad H_{-1}=i\bigl((w^+)^2\partial_++w^+y\partial_y-y^2\partial_-\bigr)\,,\\
		\bar{H}_1&=i\partial_-\,, \qquad \bar{H}_0=i\bigl(w^-\partial_-+\frac{1}{2}y\partial_y\bigr)\,, \qquad \bar{H}_{-1}=i\bigl((w^-)^2\partial_-+w^-y\partial_y-y^2\partial_+\bigr)\,.\\
	\end{split}
\end{equation}
Each line of these isometry generators satisfies the $SL(2,\mathbb{R})$ algebra \eqref{algebra} and has quadratic Casimir 
\begin{equation}\label{confcoordcas}
	\mathcal{H}^2=\frac{1}{4}(y^2\partial_y^2-y\partial_y)+y^2\partial_+\partial_-\, .
\end{equation}
Setting $L=1$ from now on, this Casimir coincides with the full scalar Laplacian on the Poincar\'e patch metric \eqref{poincare}.
Similarly, for the spacetimes that we consider, we will show that the scalar Laplacian can also be matched to the $SL(2,\mathbb{R})$ quadratic Casimir $\mathcal{H}^2$ in an expansion around $r_+$. 
Our method explicitly recovers the Love symmetry generators presented in the literature for 4-dimensional Kerr and 5-dimensional Myers--Perry spacetimes, and provides us with a means to extend the calculation of Love symmetry to higher dimensions.

The next step is to introduce a set of ``thermal'' coordinates $(t,r,\pstar)$, related to the conformal coordinates $(w^\pm,y)$ as follows:
\begin{equation}\label{conformalAnsatz}
	w^+=\sqrt{\frac{q}{q+1}}e^{\alpha\pstar+\beta t}\,,\quad w^-=\sqrt{\frac{q}{q+1}}e^{\gamma\pstar+\delta t}\,,\quad y=\frac{1}{\sqrt{q+1}}e^{\frac{1}{2}[(\alpha+\gamma)\pstar+(\beta+\delta) t]}\,,
\end{equation}
where $(\alpha, \beta, \gamma, \delta)$ are 4 independent parameters, and $q=q(r)$ is a function of the radial coordinate $r$.  
We will fix $(\alpha, \beta, \gamma, \delta)$ as well as the function $q$ for each black hole of interest. 
The coordinates $t$ and $\eta$ in \eqref{conformalAnsatz} correspond to the Killing vectors $\partial_t$ and $\partial_\eta$ defined in equation \eqref{xi}.  

Using this coordinate transformation, the generators take the following form:
\begin{equation}\label{moregengenerators}
	\begin{split}
		H_0&=\frac{i}{\beta\gamma-\alpha\delta}\left(\gamma\partial_t-\delta\partial_{\pstar}\right)\,,\\
  H_{\pm1}&=\frac{ie^{\mp(\alpha\pstar+\beta t)}}{\sqrt{q(1+q)}}\left(\pm\frac{q(1+q)}{q'}\partial_r+\frac{\left(\alpha+\gamma(1+2q)\right)\partial_t-\left(\beta+\delta(1+2q)\right)\partial_{\pstar}}{2(\beta\gamma-\alpha\delta)}\right)\,, \\
		\bar{H}_0&=\frac{-i}{\beta\gamma-\alpha\delta}\left(\alpha\partial_t-\beta\partial_{\pstar}\right)\,,\\ \bar{H}_{\pm1}&=\frac{ie^{\mp(\gamma\pstar+\delta t)}}{\sqrt{q(1+q)}} \left(\pm\frac{q(1+q)}{q'}\partial_r+\frac{-\left(\gamma+\alpha(1+2q)\right)\partial_t+\left(\delta+\beta(1+2q)\right)\partial_{\pstar}}{2(\beta\gamma-\alpha\delta)}\right)\,,\\
	\end{split}
\end{equation}
while the quadratic Casimir \eqref{confcoordcas} reads:
\begin{equation}\label{FullCas}
	\begin{split}
		\mathcal{H}^2 =& \frac{q(q+1)}{(q')^2}\partial_r^2 + \left(\frac{q+1}{q'}
        \left(\frac{q}{q'}\right)'+\frac{q}{q'}\right)\partial_r
		\\&
		-\frac{1}{q}\left(\frac{(\gamma+\alpha)\partial_t-(\beta+\delta)\partial_{\pstar}}{2(\beta\gamma-\alpha\delta)}\right)^2+\frac{1}{1+q}\left(\frac{(\gamma-\alpha)\partial_t+(\beta-\delta)\partial_{\pstar}}{2(\beta\gamma-\alpha\delta)}\right)^2		.
	\end{split}
\end{equation} 
By construction, the $H$s and $\bar{H}$s each separately satisfy the $SL(2,\mathbb{R})$ algebra \eqref{algebra}.

Here, we have used the conformal coordinates in \eqref{conformalAnsatz} merely to propose generators in thermal black hole coordinates, \eqref{moregengenerators}, which satisfy $SL(2,\mathbb{R})$ commutation relations.
We should actually think of these generators as acting on the full black hole spacetime (just somewhat un-interestingly on any directions apart from $t,\eta,r$). 
For these generators to have any meaning in the black hole spacetimes of interest, however, we wish to match their quadratic Casimir \eqref{FullCas} to the Klein--Gordon equation in the full black hole spacetime.

\subsection{Matching Radial Derivatives}
\label{sec: Radial derivs}
We begin by matching the radial derivatives of the quadratic Casimir \eqref{FullCas} to the massless Klein--Gordon equation for a scalar field $\Psi$.  
This matching will determine the function $q(r)$ introduced in the conformal coordinates \eqref{conformalAnsatz}.

Rather than start with the general form \eqref{scalar}, we will assume (radial) separability of the massless Klein--Gordon equation on the spacetime. 
That is, we assume that it is possible to find the solution of \eqref{scalar} in the form 
\be 
\Psi=R\tilde\Psi\,,
\ee
where the dependence on $r$ is entirely captured by the radial part $R$, and $\tilde \Psi$ solely depends on the remaining coordinates. Accordingly, we assume, as will be the case of the spacetimes we examine in this paper, that the radially separated equation becomes
\begin{equation}\label{KGIsolatingK}
    \frac{r^2}{g_r}\partial_r\left(\Delta\partial_rR\right)+\frac{R}{\tilde\Psi}\hat{D}(\tilde\Psi) = K R\,,
\end{equation}
where $K$ is the corresponding separation constant,%
\footnote{In particular, for the case of the  four-dimensional Kerr geometry, $K$ is related to the Carter's separation constant (denoted by  $K_1$ in the higher-dimensional notation of \cite{Frolov:2017kze, Keeler:2021tqy}).} %
and functions $\Delta=\Delta(r)$, $g_r=g_r(r),$ and the operator ${\tilde \Psi}^{-1}\hat D(\tilde \Psi)=[{\tilde \Psi}^{-1}\hat D(\tilde \Psi)](r)$ are all dependent on the radial coordinate only (while $\hat D$ acts non-trivially only on non-radial coordinates, it may be $r$-dependent).

We will allow ourselves to multiply this equation by an overall {\it constant} $s$ (not a function as considered previously in \cite{Keeler:2021tqy}) 
before matching it with the quadratic Casimir \eqref{FullCas}.%
\footnote{The proportionality factor $s$ turns out to be necessary when trying to match in the higher-dimensional spacetimes. We consider only a constant $s$ because we need an eigenvalue equation for the Casimir ${\cal H}^2\Psi=\hat{\ell}(\hat{\ell}+1)\Psi$ for some $\hat{\ell}\in\mathbb{R}$.}
Thus, ultimately we seek to replace the Klein--Gordon equation in \eqref{KGIsolatingK} by the effective equation
\begin{equation}\label{eq: eff KG}
    {\cal H}^2 \Psi = s K \Psi\,.
\end{equation}
Before we begin the matching procedure, we should note that we will not actually be able to rewrite \eqref{KGIsolatingK} in the exact form of \eqref{eq: eff KG}.  
Rather, there will be terms on the left hand side of \eqref{KGIsolatingK} that are not reproduced by ${\cal H}$; as we detail below, these terms will be kept small in a near-region (and when necessary near horizon) limit. 
We will cover these limits in more detail in Section \ref{sec:LoveKerr}.

We will now continue with our matching goal. 
Writing only the radial derivative pieces, we thus require
\begin{equation}
    s\frac{r^2}{g_r}\Delta \partial_r^2 + s\frac{r^2}{g_r}\Delta' \partial_r=\frac{q(q+1)}{(q')^2}\partial_r^2+\left(\frac{q+1}{q'}\left(\frac{q}{q'}\right)'+\frac{q}{q'}\right)\partial_r.
\end{equation}
First matching the ratio of the $\partial_r$ and $\partial_r^2$ terms, we find
\begin{equation}
    \frac{\Delta'}{\Delta}=\frac{q'(q+1)(q/q')'+qq'}{q(q+1)}.
\end{equation}
Integrating both sides and introducing the integration constant $c_1$, we find 
\begin{equation}\label{DeltaFromq}
    \Delta = 2c_1\frac{q(1+q)}{q'}.
\end{equation}
We can now easily match the $\partial_r^2$ terms, finding
\begin{equation}\label{qpFromq}
    q'=\frac{g_r}{2c_1 s r^2}.
\end{equation}

Since we have insisted that the quadratic Casimir of our generators match the Klein--Gordon equation \eqref{KGIsolatingK} up to a constant $s$ (and in an expansion near $r_+$), we can now identify
\begin{equation}
    q=\frac{1}{2c_1 s}\int \frac{g_r}{r^2}dr+c_2.
\end{equation}
Here we have explicitly written the integration constant $c_2$.
We can now also find an expression for $\Delta$, using \eqref{DeltaFromq}.  
Note this $\Delta$ arises from the quadratic Casimir of the generators $H$ or $\bar{H}$.  
As such, we will refer to it as $\Deff$ since it is only an effective $\Delta$.  It takes the form
\begin{equation}\label{eq: Deff}
    \Deff=\frac{r^2}{sg_r}\left(\int \frac{g_r}{r^2}dr + 2c_2c_1s\right)\left(\int \frac{g_r}{r^2}dr + 2c_2c_1s+2c_1s\right).
\end{equation}
Since there are three constants, we will be able to match $\Deff\approx\Delta$ only approximately.  
We will perform this matching in an expansion around $r=r_+$, thus finding $\Delta=\Deff+\mathcal{O}\bigl([r-r_+]^3\bigr).$

For the spacetimes studied in this paper, $g_r$ takes the form
\begin{equation}
    g_r = c_r r^{k+1},
\end{equation}
where $c_r$ is a constant. 
For these cases, we then find
\begin{equation}\label{DeltaEffectiveForPowergr}
    \Deff=\frac{c_r}{sk^2 r^{k-1}}\left(r^k-B\right)\left(r^k-C\right), \qquad
    q=\frac{1}{B-C}\left(r^k-B\right).
\end{equation}
Here, $B=-2sc_1c_2k/c_r$ and $C=-2sc_1(1+c_2)k/c_r$ are relabelings of the two integration constants.  
As for the generic form of $g_r$, we will be able to use the three unfixed constants $s,\, B,\, C$ to match $\Delta$ through second order around $r_+$. 
Without loss of generality, in what follows we will set $B=r_+^k$, identifying $q$ with the $r_+$ pole.%
\footnote{Actually, our generators \eqref{moregengenerators} possess the symmetry $q\rightarrow1+q$ symmetry, putting the outer and inner horizons on equal footing.}

In specific cases, such as the Kerr black hole, the five-dimensional Myers--Perry black hole, and the Einstein Lense--Thirring spacetimes in general dimension, we will find exact matching of $\Deff=\Delta$.  
Even in these cases, however, the massless Klein--Gordon operators only match perfectly in the radial derivative terms; in this sense the hidden conformal symmetry only arises in a near-region limit of the original massless scalar probe in the spacetime.

We should recall that our conformal coordinates have four other parameters to be fixed, the coefficients $\{\alpha,\, \beta,\, \gamma,\,\delta\}$ in \eqref{conformalAnsatz}.  
As we will illustrate below, different choices of these parameters match  different choices of the effective symmetry generators from the literature.  
In order to examine these choices more precisely, we will first recover the generators \eqref{moregengenerators} that were discovered for non-extremal Kerr \cite{Castro:2010fd,Charalambous:2021kcz,Hui:2022vbh} and the 5D Myers--Perry black hole \cite{Charalambous:2021mea}. 
We will then extend this formalism to higher-dimensional spacetimes, utilizing the Lense--Thirring solution as our primary example.

\subsection{Love and hidden conformal symmetry for Kerr revisited}\label{sec:LoveKerr}
Using the standard Boyer--Lindquist coordinates, the Kerr metric with mass $M$ and angular momentum $J$ reads
\be
ds^2=-\frac{\Delta}{\Sigma}(dt-a\sin^2\!\theta d\phi)^2+\frac{\Sigma}{\Delta}dr^2+\Sigma d\theta^2+\frac{\sin^2\!\theta}{\Sigma}\Bigl((r^2+a^2)d\phi-adt\Bigr)^2\,,
\ee
where
\be
\Sigma=r^2+a^2\cos^2\!\theta\,,\quad \Delta=r^2-2Mr+a^2\,,\quad a=J/M\,. 
\ee
It is well known \cite{Carter:1968ks}, that the massless Klein--Gordon equation is separable on this background, with the corresponding radial equation taking the following form:  
\begin{multline}\label{KerrKG}
\Bigg[\partial_r(\Delta\partial_r)-\frac{(r_+-r_-)}{r-r_+}\left(\frac{\partial_t+\Omega_+\partial_\phi}{2\kappa_+}\right)^2+\frac{(r_+-r_-)}{r-r_-}\left(\frac{\partial_t+\Omega_-\partial_\phi}{2\kappa_-}\right)^2
\\
-(r^2+2M(r+2M))\partial_t^2\Bigg]\Psi=K\Psi.
\end{multline}
In the above equation, $K$ is the separation constant between the $\theta$ and $r$ equations, $r_\pm$ are the inner and outer horizons, and 
\begin{equation}
    \Omega_\pm=\frac{a}{2Mr_\pm}, \qquad \kappa_\pm=\frac{r_+-r_-}{4Mr_\pm}
\end{equation}
are the angular velocities and surfaces gravities at $r_\pm$, respectively. 
The angular equation is
\begin{equation}
   \left[ \frac{1}{\sin\theta}\partial_\theta(\sin\theta\partial_\theta)+\frac{1}{\sin^2\theta}\partial_\phi^2-a^2\cos^2\theta\partial_t^2\right]\Psi=-K\Psi.
\end{equation}

Our first step in studying the hidden conformal symmetry of Kerr is to match the radial derivative terms of \eqref{KerrKG} to those of the quadratic Casimir \eqref{FullCas}.  
Comparing the Kerr separated radial equation to the generic form \eqref{KGIsolatingK}, we find $g_r=r^2$.  
We thus have $c_r=k=1$ in our expression for $q$ and $\Deff$
\eqref{DeltaEffectiveForPowergr}:
\begin{equation}\label{Deff}
    \Deff=\frac{1}{s}(r-B)(r-C).
\end{equation}
Formally, we could fix the three constants $s,\, B,$ and $C$ by expanding both $\Deff$ and the actual $\Delta$ around $r_+$.  However in this case, we can clearly match exactly by setting $s=1$, $B=r_+$, and $C=r_-$.
Making this choice gives
\begin{equation}\label{qeq}
    q(r)=\frac{r-r_+}{r_+-r_-}.
\end{equation} 

To determine the parameters $(\alpha,\beta,\gamma,\delta)$ for Kerr, we have to match more of the operator \eqref{KerrKG} and \eqref{FullCas}. Since different choices of the parameters are made in the literature, we will review some of these choices in the following subsections, and explore the physics that they imply.

\subsubsection{Locally-defined hidden conformal symmetry generators for Kerr}\label{subsec:CMS}

In this section, we review the ideology and results of \cite{Castro:2010fd}, which succeeded in constructing a \textit{locally-defined} set of $SL(2,\mathbb{R})$ generators. 
In that work, the authors were interested in matching the entire quadratic Casimir \eqref{FullCas} to the top line of the Klein--Gordon operator in \eqref{KerrKG}. 
Consequently, they discarded the non-pole terms in the radial Klein--Gordon equation by invoking a so-called near-region (or soft hair) limit%
\footnote{The near-region limit has also been called the ``near-zone'' limit \cite{Charalambous:2021kcz, Hui:2022vbh} and ``soft hair'' limit \cite{Haco:2018ske}. 
In this work we will stick with the phrase ``near-region'' limit. Many authors  \cite{Krishnan:2010pv,Haco:2018ske,Perry:2020ndy,Charalambous:2021kcz,Charalambous:2023jgq} have also employed this  limit in studying hidden conformal symmetry. 
See also \cite{Chanson:2022wls} for a discussion of black ring solutions in this context. } in the dynamics: $\omega M\ll1$ and $\omega r\ll1$.

In our language, the matching that \cite{Castro:2010fd} carried out is essentially
\begin{equation}\label{CMSsplit}
    \begin{split}
        &-\frac{1}{q}\left(\frac{(\gamma+\alpha)\partial_t-(\beta+\delta)\partial_{\eta}}{2(\beta\gamma-\alpha\delta)}\right)^2+\frac{1}{1+q}\left(\frac{(\gamma-\alpha)\partial_t+(\beta-\delta)\partial_{\eta}}{2(\beta\gamma-\alpha\delta)}\right)^2=\\
        &-\frac{(r_+-r_-)}{r-r_+}\left(\frac{\partial_t+\Omega_+\partial_\phi}{2\kappa_+}\right)^2+\frac{(r_+-r_-)}{r-r_-}\left(\frac{\partial_t+\Omega_-\partial_\phi}{2\kappa_-}\right)^2,
    \end{split}
\end{equation}
where we should identify $\partial_\eta=\Omega_+\partial_\phi$. As discussed above, the proportionality constant $s=1$ and $q$ is given in \eqref{qeq}. 
Equating the coefficients of the $\partial_\phi^2$, $\partial_\phi\partial_t$ and $\partial_t^2$ terms in each pole, we find the following conformal coordinate parameters:
\begin{equation}\label{eq: CMS params}
    \alpha=\kappa_+, \qquad \beta=0, \qquad \gamma=2\pi T_L, \qquad \delta=-\frac{1}{2M},
\end{equation}
where $T_R=\frac{r_+-r_-}{4\pi a}$ and $T_L=\frac{r_++r_-}{4\pi a}$. The resulting generators are
\begin{equation}\label{BLHs}
\begin{split}
H_{\pm 1}&=i e^{\mp2 \pi  T_R {\phi }}\bigg(\pm\Delta^{1/2}\partial_{{r}}+\frac{1}{2\pi T_R}\frac{{r}-M}{\Delta^{1/2}}\partial_{{\phi }}+\frac{2T_L}{T_R}\frac{M({r}-M)}{\Delta^{1/2}}\partial_{{t}}\bigg)\,,\\
H_0&=\frac{i}{2\pi T_R}\partial_{{\phi }}+2iM\frac{T_L}{T_R}\partial_{{t}}\,,\\
\bar{H}_{\pm 1}&=i e^{\mp(2 \pi  T_L {\phi }-\frac{{t}}{2M})}\bigg(\pm\Delta^{1/2}\partial_{{r}}-\frac{M}{\Delta^{1/2}}\partial_{{\phi }}-2M\frac{{r}}{\Delta^{1/2}}\partial_{{t}}\bigg)\,,\\
\bar{H}_0&=-2iM\partial_{{t}}\,.\\
\end{split}
\end{equation}
As discussed in \cite{Castro:2010fd}, only two of the generators \eqref{BLHs} are globally well-defined under $\phi\rightarrow\phi+2\pi$, and thus we have recovered  a local $SL(2,\mathbb{R})\times SL(2,\mathbb{R})$ hidden conformal symmetry. 

\subsubsection{Globally-defined hidden conformal symmetry generators for Kerr} 

In this section, we systematically construct globally-defined hidden conformal symmetry generators for the Kerr black hole. 
We obtain a 1-parameter family of such  generators, with our results matching those of \cite{Charalambous:2021kcz} for $\beta=0$. 
In \cite{Charalambous:2021kcz}, the authors discard a seemingly ad hoc piece of the $(r-r_-)$ pole of the radial Klein--Gordon operator (see equation \eqref{kerrdrop} with $\beta=0$), noting their alternative truncation is also consistent with the soft hair limit proposed by \cite{Castro:2010fd}.
 We hope that our method provides a ``standardized'' approach to obtaining truncations which yield globally-defined $SL(2,\mathbb{R})$ generators near the outer horizon.

Having carried out the radial derivative matching in \eqref{Deff} and \eqref{qeq}, we now match the $r_+$ pole of the quadratic Casimir \eqref{FullCas} to that of the Klein--Gordon operator \eqref{KerrKG}
\begin{equation}\label{KerrRPpole}
    \begin{split}
        &-\frac{1}{q}\left(\frac{(\gamma+\alpha)\partial_t-(\beta+\delta)\partial_{\eta}}{2(\beta\gamma-\alpha\delta)}\right)^2=-\frac{(r_+-r_-)}{r-r_+}\left(\frac{\partial_t+\Omega_+\partial_\phi}{2\kappa_+}\right)^2.
    \end{split}
\end{equation}
It is evident from equation \eqref{moregengenerators} that there are two equivalent conditions that render globally-defined symmetry generators: either one sets $\gamma=0$ or $\alpha=0$. 
Without loss of generality, we examine the $\gamma=0$ case, as this set reproduces the generators reported in \cite{Charalambous:2021kcz}. 

In setting $\gamma=0$, the matching condition \eqref{KerrRPpole} determines 3 of the 4 conformal coordinate parameters:%
\begin{equation}\label{kerrconditions}
    \gamma=0, \qquad \delta=-\sigma\kappa_+, \qquad \alpha=\sigma\kappa_+-\beta
    ,
\end{equation}
where $\sigma\equiv\pm 1$.
Recalling that $\partial_\eta = \Omega_+\partial_\phi$, these parameters yield the following globally-defined set of $SL(2,\mathbb{R})\times U(1)$ generators:
\begin{equation}\label{Kerrgens}
	\begin{split}
		H_0&=\frac{i}{\sigma\kappa_+-\beta}\partial_{\pstar}\,,\\
		\bar{H}_0&=\frac{-i}{\sigma\kappa_+}\left(\partial_t-\frac{\beta}{\sigma\kappa_+-\beta}\partial_{\pstar}\right)\,,\\
  \bar{H}_{\pm1}&=\frac{ie^{\pm \sigma\kappa_+t}}{\sqrt{\Delta}} \left(\pm\Delta\partial_r+\frac{\Delta'}{\sigma\kappa_+}\partial_t-\frac{\left(-\sigma(r_+-r_-)\kappa_++\beta\Delta'\right)}{2\sigma\kappa_+(\sigma\kappa_+-\beta)}\partial_\eta\right)\,.\\
	\end{split}
\end{equation}

The generators reported in \cite{Charalambous:2021kcz} are a special instance of \eqref{Kerrgens}, when one selects $\sigma=+1$ and sets $\beta=0$.%
\footnote{To go between the generators in \eqref{Kerrgens} and those presented in \cite{Charalambous:2021kcz}, use $\bar{H}_0=iL_0$ and $\bar{H}_{\pm 1}=-iL_{\pm 1}$, and note that their $\beta=1/\kappa_+$.} %
In that case, the conformal coordinate parameters are
\begin{equation}\label{KerrLoveconfcoord}
	\gamma=\beta=0, \qquad \alpha=2\pi T_R\equiv\frac{\kappa_+}{\Omega_+}, \qquad\delta=-\kappa_+.
\end{equation}
The globally-defined generators in \eqref{Kerrgens} thus become
\begin{equation}
	\begin{split}
		\bar{H}_0&=-i\kappa_+^{-1}\partial_t\,, ~\qquad \bar{H}_{\pm 1}=-ie^{\pm\kappa_+ t}\left(\frac{a}{\Omega_+\Delta^{1/2}}\partial_{\eta}\mp\Delta^{1/2}\partial_r+\frac{\Delta'}{2\kappa_+\Delta^{1/2}}\partial_t\right)\,, \\ H_0&=i\kappa_+^{-1}\partial_{\eta}\,,\\
	\end{split}
\end{equation}
 which indeed are the generators reported in \cite{Charalambous:2021kcz}. 

It is important to note that, although many authors focus on turning the radial equation into an $SL(2,\mathbb{R})$ Casimir, the true emergent symmetry group one finds in the dynamics is $SL(2,\mathbb{R})\times U(1)$. 
The $SL(2,\mathbb{R})\times U(1)$ Casimir is 
\begin{equation}\label{casimir II}
	\begin{split}
		\mathcal{\hat{H}}^2&=-\Bar{H}_0^2+\frac{1}{2}\left(\Bar{H}_1\Bar{H}_{-1}+\Bar{H}_{-1}\Bar{H}_1\right)+\lambda\partial_\eta^2\,,\\
	\end{split}
\end{equation}
where $\lambda$ is a constant to be determined. 
The authors of \cite{Martin:2022duk} made use of the $U(1)$ part of the Casimir to fix some of the $\pm$ branches that arise in determining $(\alpha, \beta,\gamma,\delta)$ (see, for example, \eqref{kerrconditions}). 

To see the role of the $U(1)$ piece of the Casimir in the context of Love symmetry, it is instructive to inspect the $r_-$ poles of the Casimir \eqref{FullCas} and the 
Klein--Gordon operator \eqref{KerrKG}. As already discussed, we cannot match these two poles exactly. However, we can parametrize the mismatch by writing
\begin{equation}\label{rminuscompare}
\frac{1}{1+q}\left(\frac{(\gamma-\alpha)\partial_t+(\beta-\delta)\partial_{\eta}}{2(\beta\gamma-\alpha\delta)}\right)^2=\frac{(r_+-r_-)}{r-r_-}\left(\frac{\partial_t+\Omega_-\partial_\phi}{2\kappa_-}\right)^2 -X ,
\end{equation}
where $X$ represents the unmatched piece of the $r_-$ pole. 
After some algebra, we can show that
\begin{equation}\label{kerrdrop}
X=\frac{r_+-r_-}{4\kappa_+^2(r-r_-)}\left[
\left(1-\frac{\kappa_+^2}{\kappa_-^2}\right)\partial_t^2-2\left(\frac{\sigma \kappa_+ + \beta}{\sigma \kappa_+ - \beta}+\frac{\kappa_+}{\kappa_-}\right)\partial_t\partial_\eta+\left(\left(\frac{\beta+\sigma \kappa_+}{\beta-\sigma\kappa_+}\right)^2-1\right)\partial_\eta^2
\right] .
\end{equation}

The usual near-region limit imposes that $\omega M \ll 1$ and $\omega r \ll 1$, where we act our Klein--Gordon equation on a function  $\Psi\propto e^{i(m\phi-\omega t)}$.  
In this limit, the first two terms in the bracket above vanish, but the third term does not. The authors of \cite{Charalambous:2021kcz} set $\beta=0$, such that the third term also vanishes, imposing that the Klein--Gordon equation matches the quadratic Casimir in the near-region limit.

However we take a different approach.  The $\partial_\eta^2$ term can instead be thought of as the $U(1)$ piece of the full quadratic Casimir  \eqref{casimir II}. If we set 
\begin{equation}\label{eq: lambda}
    \lambda = -\frac{\sigma\beta}{\kappa_+(\beta-\sigma\kappa_+)^2}\,,
\end{equation}
then, in the {near-region limit}, the $SL(2,\mathbb{R})$ Casimir differs from the Klein--Gordon operator on the left-hand side of \eqref{KerrKG} by 
\begin{equation}
    X\rightarrow -\frac{r_+-r_-}{r-r_-}\lambda \partial_\eta^2.
\end{equation}
However, if we additionally expand around the outer horizon $r=r_+$, then the full $SL(2,\mathbb{R})\times U(1)$ quadratic Casimir \eqref{casimir II} becomes
\begin{equation}
    \mathcal{\hat{H}}^2 = -\bar{H}_0^2+\frac{1}{2}\left(\bar{H}_1\bar{H}_{-1}+\bar{H}_{-1}\bar{H}_1\right)-\frac{\sigma \beta}{\kappa_+}H_0^2,
\end{equation}
which matches the Klein--Gordon operator entirely in the {near-region limit} expanded around $r=r_+$.  
Thus, we do not need to match the $\partial_\eta^2$ piece of the Klein--Gordon operator to the $SL(2,\mathbb{R})$ Casimir; a mismatch in that piece is consistent with the presence of the $U(1)$.

In closing, we note that the 1-parameter family of generators \eqref{Kerrgens} we derive here differs from previous families, including the set in \cite{Lowe:2011aa}.  
We impose global well-definedness and matching in the $r-r_+$ poles of the quadratic Casimir and the Klein--Gordon operator, while \cite{Lowe:2011aa} ignores global well-definedness but instead matches the $\partial_\eta^2$ piece of the $r-r_-$ pole. 
The two families only overlap when $\beta=\gamma=0$, since that is where their generators becomes globally well-defined and where ours do not require using $U(1)$ in the full quadratic Casimir.  
Incidentally, this location is precisely the generator set found by \cite{Charalambous:2021kcz}.%
\footnote{The \cite{Charalambous:2021kcz} generators can be recovered from \cite{Lowe:2011aa} when their $\kappa$ is set to 1.  
Similar but more complicated algebra recovers the generators in \cite{Hui:2022vbh} when $\kappa=-1$; that value corresponds to the choice $\alpha=0$ instead of $\gamma=0$ we used for global well-definedness here. 
We thank Adam Solomon for clarifying this point for us. 
}%

\subsection{Love generators for 5D Myers--Perry via near horizon expansion}\label{sec:MP}

Let us now turn to the 5D Myers--Perry metric, which we write in the following form:
\begin{align}
    ds^2=& -dt^2 +\frac{r \rho^2}{\Delta}dr^2 + \rho^2 d\theta^2+(r^2+a^2)\sin^2 \theta d\phi^2 
    + (r^2+b^2) \cos^2 \theta d\psi^2
   \nonumber \\
    &+\frac{2M}{\rho^2}\left[dt+a\sin^2\theta d\phi + b \cos^2\theta d\psi\right]^2\,,
\end{align}
where we have set%
\footnote{Note our $\Delta$ is $\Delta/r$ in \cite{Frolov:2002xf}.}
\begin{align}\label{Delta5DMyersPerry}
    \Delta&=\frac{(r^2-r_+^2)(r^2-r_-^2)}{r}=\frac{(r^2+a^2)(r^2+b^2)-2M r^2}{r}\,,
    \\
    \rho^2 &=r^2+a^2 \cos^2 \theta+b^2\sin^2\theta\,.
\end{align}
Here $r_\pm$ are the inner and outer horizons set by the zeros of $\Delta$, and $a,\, b$ are the black hole rotation parameters, and $M$ is the black hole mass.

The Klein--Gordon equation is known to be separable, as independently found in  \cite{Cvetic:1997uw,Frolov:2002xf}. 
Namely, we consider the following separability ansatz:
\begin{equation}
	\Psi=R(x)\Theta(\theta)e^{i\left(-\omega t+m_\phi\phi+m_\psi\psi\right)}\,,
\end{equation}
where $\phi$ and $\psi$ are the angular rotation directions, and $\Psi$ is a massless scalar.  
Its radial Klein--Gordon equation then becomes
\begin{equation}\label{Radial5DMyersPerry}
\begin{split}
   \left[ \frac{1}{r}\partial_r\left(\Delta \partial_r\right)-(r^2+a^2+b^2+2M)\partial_t^2
    -\frac{r_+^2-r_-^2}{r^2-r_+^2}\frac{1}{\kappa_+^2}\left(\partial_t + \Omega_+^\psi \partial_\psi + \Omega_+^\phi \partial_\phi\right)^2\right.
    \\
    \left.+\frac{r_+^2-r_-^2}{r^2-r_-^2}\frac{1}{\kappa_-^2}\left(\partial_t + \Omega_-^\psi \partial_\psi + \Omega_-^\phi \partial_\phi\right)^2\right]\Psi= \lambda \Psi.
\end{split}
\end{equation}
Here we have used the surface gravities $\kappa_\pm$ and angular velocities $\Omega^{\phi,\psi}_\pm$, which are given by%
\footnote{Our values match those of \cite{Castro:2013kea}, except that their $\Omega_-^L=\frac{1}{2}(\Omega_-^m-\Omega_-^k)$  appears to be off by an overall minus sign. 
We believe our expressions to be correct, since it matches both \cite{Cvetic:1997uw,Frolov:2002xf}. 
Useful identities for relating our expressions to those in \cite{Castro:2013kea,Cvetic:1997uw} include $2Mr_\pm^2=(a^2+r_\pm^2)(b^2+r_\pm^2)$ and $r_+r_-=ab$; additionally their $m_R=(m_\phi+m_\psi)/2$ while their $m_L=(m_\phi-m_\psi)/2$.} 
\begin{equation}
    \kappa_\pm = \frac{r_+^2-r_-^2}{2Mr_\pm}\,, \quad \Omega^\psi_\pm=\frac{b}{b^2+r_\pm^2}\,, \quad \Omega^\phi_\pm = \frac{a}{a^2+r_\pm^2}\,.
\end{equation}

The angular separation constant $\lambda$ arises in the $\theta$ equation, which becomes
\begin{equation}
    -\partial_\theta \left(\sin \theta \cos \theta \partial_\theta\right)\Psi=\left[\lambda+(a^2+\sin^2\theta+b^2 \cos^2\theta)\partial_t^2+\frac{1}{\sin^2\theta}\partial_\phi^2+\frac{1}{\cos^2\theta}\partial_\psi^2\right]\sin\theta \cos \theta \,\Psi\,.
\end{equation}

The hidden conformal symmetry structure for the 5D Myers-Perry black hole was first reported in \cite{Krishnan:2010pv}. 
Another useful reference is Appendix A of \cite{Castro:2013kea}. Recently, the set of globally-defined Love generators for this background were presented in \cite{Charalambous:2023jgq}. 
In this section, we will show that the results of \cite{Charalambous:2023jgq} can be produced by a simple conformal coordinates argument, paving the way for obtaining Love symmetry generators for rotating black holes in general dimensions. We will demonstrate the general dimension analysis in Section \ref{sec: LT} via the example of Lense--Thirring spacetimes.%
\footnote{Our procedure is also applicable to Myers--Perry black holes in higher dimensions, although for a limited case (only one of the Killing tensor conserved quantities can be nonzero). 
We hope to return to this issue in future work.}  

The radial derivative piece of our Klein--Gordon equation \eqref{Radial5DMyersPerry} matches the general form \eqref{KGIsolatingK}, with $\Delta$ as in \eqref{Delta5DMyersPerry} and $g_r=r^3$. 
Accordingly, following our argument from Section \ref{sec: Radial derivs}, we can pick conformal coordinates whose corresponding quadratic Casimir matches the radial derivative pieces, except with
\begin{equation}
    \Deff = \frac{(r^2-B)(r^2-C)}{4sr}\,.
\end{equation}
Here, $s,\, B,\, C$ are unfixed constants.  
Just as in the four-dimensional case, we can match our actual $\Delta$ exactly, by choosing
\begin{equation}
    s= \frac{1}{4}\,, \quad B=r_+^2\,,\quad C=r_-^2\,.
\end{equation}
The conformal coordinates $w_\pm,\, y$ will then be related to $t,\, r, \eta$ via \eqref{conformalAnsatz}, with
\begin{equation}\label{q5D}
    q=\frac{r^2-r_+^2}{r_+^2-r_-^2}\,.
\end{equation}

We also need to identify the Killing direction $\partial_\eta$. Most authors choose to set one of the azimuthal quantum numbers to zero, so either $m_\phi=0$ or $m_\psi=0$ (see, for example, \cite{Krishnan:2010pv,Chanson:2022wls}); as a result, they choose either $\partial_\phi$ or $\partial_\psi$ as $\partial_\eta$.%
\footnote{This choice to turn off one of the azimuthal quantum numbers is different from considering singly-spinning spacetimes. } %
This choice comes with the interpretation that each azimuthal sector $(t,r,\phi)$ or $(t,r,\psi)$ enjoys its own distinct hidden conformal symmetry. 
However, in some non-separable cases, such as 5D black rings \cite{Emparan:2006mm}, it is not always consistent to set both azimuthal quantum numbers to zero \cite{Chanson:2022wls}. 
In such cases, only certain azimuthal sectors possess hidden conformal symmetry. 

Here, we will adopt a slightly more general approach, instead studying a conformal sector associated with a linear combination of the azimuthal Killing directions. 
We identify the angular direction $\eta$ via
\begin{equation}\label{partialEta5D}
    \partial_\eta = \Omega^\psi_+\partial_\psi + \Omega^\phi_+\partial_\phi\,,
\end{equation}
aligning with the angular operator at the outer horizon.  With this assignment, we will be able to match the entirety of the numerator of the $r_+^2$ pole term to the $1/q$ term in the quadratic Casimir \eqref{FullCas}, by appropriate choice of the constants $\alpha,\, \beta,\, \gamma,\, \delta.$

As in the 4-dimensional case, we can obtain a set of globally well-defined generators by setting $\gamma=0$ in \eqref{moregengenerators}. Matching the $r_+^2$ pole, we find
\begin{equation}\label{mpconditions}
    \gamma=0\,, \qquad \delta=-\sigma\kappa_+\,, \qquad \alpha=-\beta+\sigma\kappa_+\,.
\end{equation}
Here, $\sigma=\pm 1$ is a free sign choice. 
This set of parameters is very similar to the Kerr case \eqref{kerrconditions}, when additionally replacing $\Omega_+\partial_\phi$ from Kerr with $\partial_\eta$ for the 5D Myers--Perry as in \eqref{partialEta5D}.
The generators still look slightly different for 5D, because they also depend on $q$, which differs between the 5D Myers--Perry case \eqref{q5D} and the 4D Kerr case \eqref{qeq}.

With these assignments, the globally defined symmetry generators for Myers--Perry become%
\footnote{Setting $\beta=0$, we find that the barred generators match those reported in equation (5.6) of \cite{Charalambous:2023jgq}. }
\begin{equation}\label{Mpgens}
    \begin{split}
    H_0&=\frac{i}{- \beta + \sigma\kappa_+}\partial_\eta\,,\\
    \bar{H}_0 &= \frac{i}{-\sigma\kappa_+}\partial_t-\frac{i\beta}{\sigma \beta \kappa_+-\sigma \kappa_+^2}\partial_\eta\,,\\
    \bar{H}_{\pm 1}&=i e^{\pm \sigma\kappa_+ t}\Bigl[\pm\left(\frac{(r_+^2-r_-^2)^2}{2r\sqrt{r\Delta}}\right)\partial_r-\left(\frac{\partial_r(\sqrt{r\Delta})}{2r\sigma\kappa_+}\right)\left(\partial_t-\frac{\beta}{-\beta+\sigma\kappa_+}\partial_\eta\right)\\
    &\qquad\qquad\qquad-\left(\frac{r_+^2-r_-^2}{2(-\beta+\sigma\kappa_+)\sqrt{r\Delta}}\right)\partial_\eta\Bigr]\,.
    \end{split}
\end{equation}
The barred generators form the $SL(2,\mathbb{R})$, while the $H_0$ forms an independent $U(1)$. 

As in the four-dimensional case when we insisted on globally well-defined generators, we will not be able to match the entirety of the $r_-^2$ pole term.  
However the problem here is worse; we can match only the coefficient of $\partial_\eta$, even if we ignore global well-definedness, because the numerator of the $r_-^2$ term is not just a linear combination of $\partial_t$ and $\partial_\eta$.  
Rather, defining
\begin{equation}
    \partial_{\bar{\eta}}= \Omega^\phi_+ \partial_\psi -\Omega^\psi_+ \partial_\phi\,,
\end{equation}
the numerator of the $(r^2-r_-^2)$ term in the Klein--Gordon equation \eqref{Radial5DMyersPerry} becomes
\begin{equation}
    \partial_t + \Omega_-^\psi \partial_\psi + \Omega ^\phi_- \partial_\phi = \partial_t + \frac{4M\partial_\eta + \frac{(a^2-b^2)(r_+^2-r_-^2)}{ab}\partial_{\bar{\eta}}}{ab\left(2M(a^2+b^2)-(a^2-b^2)^2\right)}\,.
\end{equation}
This $r^2-r_-^2$ term corresponds to the $1/(1+q)$ term in the $SL(2,\mathbb{R})$ quadratic Casimir \eqref{FullCas}.  
However, that quadratic Casimir only contains $\partial_t$ and $\partial_\eta$ terms, so it would be unable to match any $\partial_{\bar\eta}$ coefficients, which would block us from matching the entirety of the $r_-^2$ pole term. 
Similar difficulties arose in previous works, but were resolved in different ways. 
References \cite{Krishnan:2010pv} and \cite{Castro:2013kea} turned on only one angular momentum at a time, setting either $\partial_\phi$ or $\partial_\psi$ to zero. 
Our work more naturally matches to \cite{Charalambous:2023jgq}, which proposed a set of globally-defined Love generators for this background.

Instead, we will again consider an extended quadratic Casimir as in the Kerr case.  
Here, we have the $SL(2,\mathbb{R})\times U(1)$ generators \eqref{Mpgens}. 
In addition, since the $\psi$ and $\phi$ are the azimuthal angles, the other linear combination $\partial_{\bar{\eta}}$ also commutes with all of the $\bar{H}$ as well as the $H_0$ generator, giving a full algebra of $SL(2,\mathbb{R})\times U(1)^2$. 
Again, as in the 4-dimensional case, we can use this larger symmetry group to build a quadratic Casimir which will match the Klein--Gordon equation in the near-region limit, as long as we also set $r\rightarrow r_+$.  
Our initial mismatch in the $r^2-r_-^2$ pole can be remedied near the horizon by adding appropriate $\partial_\eta^2$ and $\partial_{\bar\eta}^2$ terms to the quadratic Casimir.

\section{Lense--Thirring black holes}\label{sec: LT}
In this section we apply our methods to a new example of the recently proposed generalized Lense--Thirring metrics~\cite{Baines:2021qaw,Gray:2021toe,Gray:2021roq}. 
These spacetimes represent a large class of slowly rotating black holes for which we show how the hidden Love (conformal) symmetries emerge. 
We will focus on two particular examples: i) the case of Einstein gravity --- meaning slowly rotating Myers--Perry black holes in general dimensions and ii) the case of Einstein--Maxwell gravity, i.e. slowly rotating charged black holes in all dimensions greater than 3. 
We briefly discuss how to the extend the Love symmetry construction to generic black holes within this Lense--Thirring class.
To this end we present these Lense--Thirring metrics and describe some of their properties.

New to this work, we take inspiration from~\cite{Sadeghian:2022ihp} to put these metrics into the standard form for separation of variables and then explicitly separate the Klein--Gordon equation. 
This allows us, {using minimal assumptions stated below,} to see how the near horizon hidden conformal symmetry arises by applying our methods to the resulting radial equation.

\subsection{Spacetime properties}\label{subsec: properties}

We now introduce the generalized Lense--Thirring spacetimes and mention their hidden symmtery properties.  
The metric of these multiply-spinning spacetimes reads~\cite{Gray:2021toe,Gray:2021roq}
\be\label{LTHDimprovedApp}
ds^2=-Nfdt^2+\frac{dr^2}{f}+r^2\sum_{i=1}^m \mu_i^2\Bigl(d\phi_i+p_i dt\Bigr)^2+r^2\sum_{i=1}^{m+\epsilon}d\mu_i^2\;.
\ee
Here, $m=[\frac{d-1}{2}]$,%
\footnote{Note that, here we have introduced the standard notation $m$ as a way of parameterizing the dimensions; it should not be confused with the eigenvalue of the angular Killing vector $\partial_\phi$ in the previous sections. 
When we need the equivalent higher-dimensional constants we will use a subscript e.g. $m_i$ corresponding to the eigenvalue of $\partial_{\phi_i}$.
}
and $\epsilon=1, 0$ for even, odd dimensions $d=2m+\epsilon+1$ respectively. 
The functions $f, N, p_i$ are functions of the radial coordinate $r$, and the coordinates $\mu_i$ obey the following constraint: 
\be\label{contraint}
\sum_{i=1}^{m+\epsilon} \mu_i^2=1\,.
\ee
In \cite{Gray:2021roq}, it was useful to define 
\be
p_i=\frac{{\sum_{j=1}^mp_{ij}a_j}}{r^2}\,, 
\ee
to explicitly introduce the corresponding rotation parameters $a_i$.

Many slowly rotating spacetimes (to linear order in rotation parameters) can be written in this generalized form (by completing the square in the $t-\phi_i$ sector). 
This includes the Kerr--(A)ds black holes, Kerr--Sen black hole~\cite{Sen:1992ua}, or the $d=5$ minimal supergravity black hole of Chong--Cveti\v{c}--L\"u--Pope~\cite{Chong:2005hr}. 
However, this is only true to linear order in $a_i$. It remains a question if one can find the appropriate energy momentum tensor and concrete functions $f, N, p_i$ for this to be an exact solution of Einstein equations or their generalization to higher curvature theories discussed in~\cite{Gray:2021roq}.   

When we have a horizon (which we take to be non-degenerate and located at $r_+$ -- the largest root of $f(r)=0$), it is a Killing horizon generated by the following Killing vector:
\be\label{killinggen}
\xi=\partial_t+\sum_{i=1}^m \Omega^i_+ \partial_{\phi_i}:=\partial_t +\partial_\pstar\,,
\quad 
\Omega^i_+= -p_i(r_+)\,.
\ee
The horizon is surrounded by an ergoregion, inside of which the  Killing vector $\partial_t$ has negative norm. 
Due to this ergoregion, the metric will exhibit superradiant phenomena  \cite{Brito:2015oca}.

We also briefly mention that the metric admits a rapidly growing tower of Killing tensors which underlie the separability discussed in the next section. 
We refer the interested reader to Appendix~\ref{Appendix: LT} for full details. 
In this appendix we also demonstrate that the spacetime is in the ``standard separable form''~\cite{Benenti:1979} which underlies the separability of the geodesic and scalar wave equations.

\subsection{Separability of scalar wave equation}\label{subsec: LT Sep}

In order to separate the scalar wave equation, let us first diagonalize the $\mu_i$ sector of the metric. Following \cite{Sadeghian:2022ihp} we introduce new coordinates $x_\mu$, $\mu=1,\dots m+\epsilon-1$:
\be\label{eq:mueq}
\mu_\mu=x_\mu\sqrt{U_\mu}\,,\quad  \mu_{m+\epsilon}=\sqrt{U_{m+\epsilon}}\,,
\ee
where we have defined,
\ba\label{eq:Ueq1}
X_\mu&=&1-x_\mu^2\,,\quad 
U_{\mu}=X_1\dots X_{\mu-1}\,. 
\ea
Then we find that the constraint \eqref{contraint} is automatically satisfied, and 
\be
\sum_{i=1}^{m+\epsilon} d\mu_i^2= \sum_{\mu=1}^{m+\epsilon-1}\frac{U_\mu}{X_\mu} dx_\mu^2\,.
\ee
The metric \eqref{LTHDimprovedApp} thus reads 
\be\label{metric}
ds^2=-Nfdt^2+\frac{dr^2}{f}+r^2\sum_{i=1}^m \mu_i^2\Bigl(d\phi_i+p_i dt\Bigr)^2+r^2\sum_{\mu=1}^{m+\epsilon-1}\frac{U_\mu}{X_\mu} dx_\mu^2\,,
\ee
and has the following inverse:
\be
g^{-1}=-\frac{1}{Nf}\bigl(\partial_t-\sum_ip_i\partial_{\phi_i}\bigr)^2+f(\partial_r)^2+\frac{1}{r^2}\left[\sum_{i=1}^m
\frac{1}{\mu_i^2}(\partial_{\phi_i})^2+\sum_{\mu=1}^{m+\epsilon-1}\frac{X_\mu}{ U_\mu}(\partial_{x_\mu})^2\right]\,.
\ee
Note that the last two terms (in square brackets) of this equation are simply the inverse metric on the $(d-2)$-sphere, which we denote $\gamma^{I J}$, where $I,J$ label the coordinates $(x_\mu,\phi_i)$. 
Moreover, the determinant of the metric reads 
\be
\sqrt{-g}=\sqrt{N}r^{2m+\epsilon-1}\frac{\mu_1\dots \mu_m\sqrt{U_1\dots U_{m+\epsilon}}}{U_{m+\epsilon}}\,, 
\ee
which evidently splits into a product of functions of $r$ and $x_\mu$:
\be\label{eq: DetMet}
\sqrt{-g}\equiv g_r(r) \sqrt{\gamma}= g_r(r) g_1(x_1)\dots g_\mu(x_{m+\epsilon-1})\,, 
\ee
{where $g_r=\sqrt{N}r^{2m+\epsilon-1}=\sqrt{N}r^{d-2}$ and $g_\mu(x_\mu)=x_\mu X_\mu^{m+\epsilon/2-1-\mu}$. 
It is this product form which is crucial to the separation of variables~\cite{benenti:2002a,benenti:2002b} see Appendix~\ref{Appendix: LT 2} for more details.

Let us now consider the {massless} scalar wave equation
\be
\frac{1}{\sqrt{-g}}\Bigl(\sqrt{-g}g^{ab}\partial_b\Psi)_{,a}=0 \,, 
\ee
and use the form of the inverse metric and determinant to write
\begin{multline}\label{eq: Sep LT first step}
\frac{1}{g_r}\Bigl(g_r g^{rr}\partial_r\Psi\Bigl)_{,r}+\Bigl(g^{tt}\partial_{t}\Psi\Bigr)_{,t}+\sum_i\left[\Bigl(g^{\phi_it}\partial_{t}\Psi\Bigr)_{,\phi_i}+\Bigl(g^{t \phi_i}\partial_{\phi_i}\Psi\Bigr)_{,t}\right]+\sum_{i,j}\Bigl(g^{\phi_i \phi_j}\partial_{\phi_i}\Psi\Bigr)_{,\phi_j}
\\ +\frac{1}{r^2\sqrt{\gamma}}\Bigl(\sqrt{\gamma}\gamma^{I J}\partial_I\Psi\Bigr)_{,J}=0\,,
\end{multline}
where the last term is the Laplacian on the $(d-2)$-dimensional sphere. Therefore, when we make the  separability ansatz
\be
\Psi=R(r)\underbrace{S_1(x_1)\dots S_{m+\epsilon-1}(x_{m+\epsilon-1})e^{im_1\phi_1+\dots im_m \phi_m}}_{\equiv {\cal S}(x_\mu,\phi_i)} e^{-i\omega t}\,,
\ee
the radial separation constant here is exactly the usual spherical harmonic one.
This follows by using explicitly the angular equation on the $d-2$ sphere
\begin{align}\label{eq: partial separated}
\frac{1}{\sqrt{\gamma} }\Bigl(\sqrt{\gamma}\gamma^{I J}\partial_I{\cal S}\Bigl)_{,J}\equiv K{\cal S}
\,,\quad
 K=\ell(\ell+d-3)\,,
\end{align}
for integer $\ell$. 
Hence, the first terms in \eqref{eq: Sep LT first step} give the the radial equation (dividing by $\Psi$, multiplying by $r^2$, and substituting $K$)
\be\label{radial}
\frac{r^2}{g_r R}(f g_r\partial_r R)_{,r}+\frac{r^2}{Nf}\bigl(\omega+\sum_i p_im_i\bigr)^2-K=0\,.
\ee
The massive equation is separated similarly.

For completeness we note the angular equations separate iteratively. 
First, we substitute the definition of $\mu_{i}$ and $U_\mu$ (i.e. \eqref{eq:mueq} and \eqref{eq:Ueq1} respectively) to obtain
\begin{align}\label{eq: Spher Harm}
0&=K-\sum_{\mu=1}^{m}\frac{m_\mu^2}{\mu_\mu^2}
+\sum_{\mu=1}^{m-1+\epsilon}\frac{1}{U_\mu g_\mu S_\mu}\Bigl({X_\mu g_\mu}S'_\mu\Bigr)'\,\nonumber\\
&=K-\sum_{\mu=1}^{m-1}\frac{1}{U_{\mu}}\left[\frac{m_\mu^2}{x_\mu^2}
-\frac{1}{g_\mu S_\mu}\Bigl({X_\mu g_\mu}S'_\mu\Bigr)'\right] -(1-\epsilon)\frac{m_{m+\epsilon}^2}{U_{m+\epsilon}}\,.
\end{align}
Then we multiply by $X_1$. 
This gives an equation which only depends on $x_1$ while the other terms are independent so must equal a constant $K_1$. 
Thus we introduce a new separation constant $K_1$ and repeat the process e.g. next multiplying by $X_2$. 
The end result is the equations,
\be
\frac{1}{S_\mu g_\mu}\Bigl({X_\mu g_\mu}S'_\mu\Bigr)'-\frac{m_\mu^2}{x_\mu^2} +K_{\mu-1}= \frac{ K_\mu }{X_\mu}\,.
\ee
It is characterized by $(m+\epsilon+1)$ separation constants $K_\mu$, 
where $K_{0}\equiv K$ and $K_{m+\epsilon}=(1-\epsilon)m^2_{m+\epsilon}$.

Before we discuss the emergence of hidden conformal symmetries we need to understand the assumptions and approximations needed. 
We do this in the next section.

\subsection{Asymptotic Charges and Approximation Regions}\label{sec: Charges}
We have so far considered general functions $f, N, p_i$ with the minimal assumption that there is a non-degenerate horizon. 
Moreover, we now assume that $N$ and $p_i$ are regular for $r\geq r_+$ and that any other roots of $f$ (corresponding e.g. to an inner horizon or cosmological horizon) are far enough away from $r_+$.

{To further characterize these Lense--Thirring spacetimes we note that in the case of (vacuum) Einstein gravity the metric functions take the form
\begin{equation}\label{eq: Met func Eins}
    N^E(r)=1\,,\quad p^E_i(r)=\frac{a_i}{r^2}(f^E(r)-1)\,,\quad
    f^E(r)=1-\frac{16\pi M}{(d-2)\Omega_{d-2}r^{d-3}}=1-\left(\frac{r_+}{r}\right)^{d-3}\,.
\end{equation}
In particular, this corresponds to a slight modification of the slow rotation limit of the $d$-dimensional Myers--Perry black holes which solves the vacuum Einstein equations to linear order in the rotation parameters $a_i$. 
Thus, the asymptotic mass and angular momenta charges (calculated e.g. as ADM charges~\cite{Arnowitt:1961zz}) are
\begin{equation}
{\cal Q}(\partial_t)=M\,,\quad {\cal Q}(\partial_{\phi_i})=a_iM\equiv J_i\,.
\end{equation}
We note that the mass $M$ has dimensions of $[M]=(d-3)$ where $[r]=1$. More generally we can expect the asymptotic expansions of generic metric functions to match this form at leading order in larger $r$, see for example the expansions in~\cite{Gray:2021roq}. 
Thus we assume the spacetime to have these asymptotic charges and, in particular, asymptotic mass $M$. 

In order to match the Casimir in these Lense--Thirring spacetimes, we need to implement a version of the soft hair limit mentioned above where now, due to the higher $d$ dimensional analysis, the appropriate combination to consider small is 
$M\omega^{d-3} \ll 1$, or equivalently $\omega r_+\ll1$,  and $r\omega \ll 1$.
In the next section we carry this procedure out for Einstein and Einstein--Maxwell gravity and comment on how to generalize the procedure to the general class of these spacetimes.

\subsection{Love symmetry of Einstein Lense--Thirring metrics in general dimension}\label{subsec: Love LT}

We now work towards seeing the emergence of hidden conformal (Love) symmetry in a general-dimensional setting, we will focus here only on the set of globally well defined generators. 
To build the effective Laplacian we start by matching the radial derivative terms. 
By construction the radial equation \eqref{radial} is in exactly the same form as \eqref{KGIsolatingK}. 
So to match the derivative terms of the Casimir \eqref{FullCas} we simply need to follow the steps of section~\ref{sec: Radial derivs}. 
Identifying the function $\Delta\equiv f g_r=r^{d-2}\sqrt{N(r)} f(r)$ we can write \eqref{radial} as
\be\label{eq: LT Rad eq}
\frac{r^2}{g_r}(\Delta \partial_r\Psi)_{,r}-\frac{r^{d}}{\sqrt{N}\Delta}\bigl(\partial_t-\sum_i p_i\partial_{\phi_i}\bigr)^2\Psi-K\Psi=0\,. 
\ee
In particular this means (assuming $s$ is constant) the best we can match the radial derivatives is
\begin{align}
   \Delta&\simeq\Deff+\dots \\
   \Deff&\equiv\frac{r^{4-d}}{s\sqrt{N(r)}}
   \left(\int {r^{d-4}}{\sqrt{N}}dr - B\right)
   \left(\int {r^{d-4}} {\sqrt{N}}dr - C\right).
\end{align}
Here the dots stand for subleading terms which appear in powers of $(r-r_+)$.

In order to progress we shall focus on the simplest case of these Lense--Thirring spacetimes: that of Einstein Gravity. However, we stress that the procedure could be applied to any particular theory of interest.

\subsubsection{Einstein Lense--Thirring spacetimes}

Specifying to the Einstein solution means that the metric functions take the form of \eqref{eq: Met func Eins} and consequently
\begin{equation}
    \Delta^E=g_r^E f^E=r^{d-2}\left[1-\left(\frac{r_+}{r}\right)^{d-3}\right]=\frac{1}{r^{d-4}}\left(r^{d-3}-r_+^{d-3}\right) r^{d-3}\,.
\end{equation}
Applying the results of \eqref{sec: Radial derivs} (where in the notation of that section $c_r=1$, $k=d-3$) and using \eqref{DeltaEffectiveForPowergr} we have the \emph{exact} matching of the derivative terms. 
That is, $\Delta^E=\Deff$ for the following constants
\begin{equation}\label{eq: const sol}
  s=\frac{1}{(d-3)^2}\,,\quad B=r_+^{d-3}=\frac{16\pi M}{(d-2)\Omega_{d-2}}\,,\quad C=0\,,
\end{equation}
and 
\begin{equation}
    q(r)=\frac{1}{r_+^{d-3}}\left(r^{d-3}-r_+^{d-3} \right)\,.
\end{equation}

We now need to match the non-derivative pieces, in the near-region. 
It will be useful to introduce the dimensionless variable $\tilde{r}=r/r_+$, such that 
\begin{equation}
    q=\tilde{r}^{d-3}-1\,.
\end{equation}
So using \eqref{DeltaFromq} and \eqref{qpFromq} and the Einstein gravity expressions \eqref{eq: const sol} we find
\begin{equation}\label{eq: Deff Ein}
    \Delta_{\text{eff}}=\Delta=\frac{r_+^{2(d-3)}}{r^{d-4}}q(q+1)=r_+^{d-2}\,\tilde{r}^{4-d} q(q+1)\,.
\end{equation}

Recalling that $p_i^E=-a_i(1-f^E(r))/r^2=a_ir_+^{d-3}/r^{d-1}$, the non derivative piece then takes the form
\begin{align}\label{eq: Non deriv match 1}
    \frac{sr^{d}}{\Delta^E}\bigl(\partial_t-\sum_i p^E_i\partial_{\phi_i}\bigr)^2
    =&\frac{sr^{d}}{\Delta_{\text{eff}}}\bigl(\partial_t-\sum_i p^E_i\partial_{\phi_i}\bigr)^2
    \nonumber\\
    =&\frac{sr_+^2\,\tilde{r}^{2(d-2)}}{q(q+1)}\left(\partial_t+\sum_{i} \frac{a_i r_+^{d-3}}{r^{d-1}} \partial_{\phi_i}\right)^2
    \nonumber\\
    =&\frac{1}{(\tilde{r}^{d-3}-1)\tilde{r}^{d-1}}\left(\frac{r_+}{d-3}\left[\tilde{r}^{d-1}\partial_t+\partial_\pstar\right]\right)^2
    \nonumber\\
    =&\frac{1}{(\tilde{r}^{d-3}-1)}\left(\frac{r_+^2}{(d-3)^2}\tilde{r}^{d-1}\partial_t^2+\frac{2r_+^2}{(d-3)^2}\partial_t\partial_\pstar\right)
    \nonumber\\
    &+\frac{1}{(\tilde{r}^{d-3}-1)\tilde{r}^{d-1}}\frac{r_+^2}{(d-3)^2}\partial_\pstar^2
    \,,
\end{align}
where, recall that we have defined the Killing vector 
$\partial_\eta=\sum_{i}\Omega^i_+ \partial_{\phi_i}$ with $\Omega^i_+=-p_i(r_+)=a_i/r_+^2$ being the angular velocities along each rotation at the horizon. 
It is important to note that \eqref{eq: Non deriv match 1} has a different pole structure to the previous examples discussed. 
This is because the metric function \eqref{eq: Met func Eins} in the Einstein--Lense--Thirring case has no inner horizon, meaning that the second pole occurs at $r=0$. 
So to proceed we carefully analyse the residues appearing with the ultimate goal being, as before, to exactly match the $r_+$ pole completely, and then understand the remaining terms.

Let us do this piece by piece, expanding \eqref{eq: Non deriv match 1}. First, the $\partial_t^2$ piece takes the following form,
\begin{align}
    \frac{\tilde{r}^{d-1}}{\tilde{r}^{d-3} -1}\frac{r_+^2 \partial_t^2}{(d-3)^2}
    =&\left(\frac{1}{\tilde{r}^{d-3} -1}
    +\frac{\tilde{r}^{d-1}-1}{\tilde{r}^{d-3}-1}\right)
    \frac{r_+^2\partial_t^2}{(d-3)^2}
    \nonumber\\
    =&\frac{1}{\tilde{r}^{d-3} -1} \frac{1}{4\kappa_+^2}\partial_t^2+
    \left[1+\frac{\tilde{r}^{d-3}(1+\tilde{r})}{1+\sum_{k=1}^{d-4}\tilde{r}^k }\right]
    \frac{\partial_t^2}{4\kappa_+^2}\,,
\end{align}
where we have introduced the surface gravity $\kappa_+\equiv1/2f'(r_+)=(d-3)/(2r_+)$.%
\footnote{In going from the first to the second line we have used the identity 
$$(1-x^n)=(1-x)\sum_{k=0}^{n-1}x^k \,.$$
}
The term in square brackets is finite in the limit $r\to r_+$ ($\tilde{r}\to1$) and will be small on the solution space of $\Phi$ when $\omega r_+\ll1$, i.e. the near-region limit. 
Next the $\partial_t\partial_\pstar$ piece is already of the correct form, since the $\tilde{r}^{d-1}$ terms cancel out and we are left with just the residue. 

Finally, the $\partial_\pstar^2$ piece is,
\begin{align}
    \frac{1}{(\tilde{r}^{d-3} -1)\tilde{r}^{d-1}}
    \frac{\partial_\pstar^2}{4\kappa_+^2}
    =&
    -\left(\tilde{r}^{1-d}+\frac{1}{\tilde{r}^2} +\tilde{r}^{d-5}-\frac{\tilde{r}^{2 (d-4)}}{\tilde{r}^{d-3}-1}\right)
    \frac{\partial_\pstar^2}{4\kappa_+^2}
    \nonumber\\
    =& -\left(\tilde{r}^{1-d}+\frac{1}{\tilde{r}^2} +\tilde{r}^{d-5}\right)\frac{\partial_\pstar^2}{4\kappa_+^2}
    +\left(\frac{1}{\tilde{r}^{d-3}-1}\right)
    \frac{\partial_\pstar^2}{4\kappa_+^2}
    +\frac{\tilde{r}^{2 (d-4)}-1}{\tilde{r}^{d-3}-1} \frac{\partial_\pstar^2}{4\kappa_+^2}
    \nonumber\\
     =& -\left(\tilde{r}^{1-d}+\frac{1}{\tilde{r}^2} +\tilde{r}^{d-5}\right)\frac{\partial_\pstar^2}{4\kappa_+^2}
    +\left(\frac{1}{\tilde{r}^{d-3}-1}\right)
    \frac{\partial_\pstar^2}{4\kappa_+^2}
    +\frac{1+\sum_{k=1}^{2d-9}\tilde{r}^k}{1+\sum_{k=1}^{d-4}\tilde{r}^k } \frac{\partial_\pstar^2}{4\kappa_+^2}
    \,.
\end{align}
The first term represents the residue at $\tilde{r}=0$, notice that for $d=4$ all of the powers of $\tilde{r}$ diverge, while for $d\geq 5$ only the first two do.
The second term diverges at $\tilde{r}=1$, i.e. at the horizon $r=r_+$. Finally, the last term is finite for all real non-negative values of $\tilde{r}$, i.e. all relevant locations. 
Putting this all together, the non derivative terms can be written
\begin{align}\label{eq: Non deriv match 2}
    \frac{sr^{d}}{\Delta^E}\bigl(\partial_t-\sum_i p^E_i\partial_{\phi_i}\bigr)^2
    =&\frac{1}{\tilde{r}^{d-3}-1}\left(\frac{1}{2\kappa_+}[\partial_t+\partial_\pstar]\right)^2-\left(\frac{1}{\tilde{r}^{d-1}}+\frac{1}{\tilde{r}^2} +\tilde{r}^{d-5}\right)\frac{\partial_\pstar^2}{4\kappa_+^2}
    \nonumber\\
    &+\left[1+\frac{\tilde{r}^{d-3}(1+\tilde{r})}{1+\sum_{k=1}^{d-4}\tilde{r}^k }\right]
    \frac{\partial_t^2}{4\kappa_+^2} +\frac{1+\sum_{k=1}^{2d-9}\tilde{r}^k}{1+\sum_{k=1}^{d-4}\tilde{r}^k} \frac{\partial_\pstar^2}{4\kappa_+^2}\,.
\end{align}

We now focus on matching the $r_+$ pole exactly. 
This pole is captured by the first term in \eqref{eq: Non deriv match 2} which has the same factor of $1/q=1/(\tilde{r}^{d-3}-1)$ as our generators give rise to in the Casimir \eqref{FullCas}. 
Thus we have to equate 
\begin{equation}
    \left[\left( \frac{1}{2\kappa_+}\bigl(\partial_t+\partial_\eta\bigr)\right)^2 \right]
    =\left(\frac{(\gamma+\alpha)\partial_t-(\beta+\delta)\partial_{\eta}}{2(\beta\gamma-\alpha\delta)}\right)^2\,.
\end{equation}
As discussed previously, we know from equation \eqref{moregengenerators} that in order to obtain a set of globally-defined generators, we can set either $\gamma=0$ or $\alpha=0$ (but not both, since we want the coordinate transformation \eqref{conformalAnsatz} to be well-defined). Without loss of generality, we consider here the $\gamma=0$ case. 

Carrying out this matching, we find that our conformal coordinate parameters for the generalized Lense--Thirring spacetime
\begin{equation}\label{einltconditions}
 \qquad \delta=-\sigma\kappa_+, \qquad \alpha=\sigma\kappa_+ -\beta\,,
\end{equation}
where, as before, $\sigma=\pm1$. 
These are directly analogous to the Kerr values \eqref{kerrconditions} and the Myers--Perry values \eqref{mpconditions} and are consistent with a slow rotation generalisation of the higher dimensional Schwarzschild--Tangherlini case~\cite{Bertini:2011ga}.

Finally, we need to consider the leftover terms and check that they are equal to some $\lambda\partial_\eta^2$ (i.e. the Casimir of the $U(1)$ symmetry) plus terms which are small in the near-region $r\omega\ll1$, $r_+\omega\ll1$ \emph{and} near the horizon. 
The remaining difference between the quadratic Casimir and the Klein Gordon equation is given by
\begin{multline}\label{eq: q=-1 pole matching}
   -\left(\frac{d-1}{d-3}\right)\frac{\partial_t^2-\partial_\eta^2}{4k_+^2} + {\cal O}(r-r_+)\\
     -\frac{(\partial_t (\beta -\kappa_+ \sigma )+\partial_\eta (\beta +\kappa_+ \sigma ))^2}{4 \kappa_+^2 (\beta -\kappa_+ \sigma )^2}
     =\lambda \partial_\eta^2 + {\cal O}(r-r_+) + {\cal O}(r_+\omega)\,,
\end{multline}
where we have expanded the unmatched terms in the non-derivative piece \eqref{eq: Non deriv match 2} around $r_+$. 
In the {near-region} we can additionally drop the $\partial_t$ terms as these terms generate $\omega r_+$ terms when acting on $\Psi$. 
This leaves only terms proportional to $\partial_\pstar^2$, thus we find for $\lambda$
\begin{equation}\label{eq: lambda LT}
     \lambda=-\frac{\sigma\beta}{\kappa_+(\beta-\sigma\kappa_+)^2}+\frac{1}{2\kappa_+^2}\frac{1}{(d-3)}
\end{equation}
which differs from the value of $\lambda$ obtained previously \eqref{eq: lambda} by the last term.

Finally, the Love generators are then (cf. \eqref{moregengenerators})
\begin{equation}\label{eq: LT generators}
	\begin{split}
        H_0&=\frac{i}{- \beta + \sigma\kappa_+}\partial_\eta\\
      \bar{H}_0&=\frac{-i}{\sigma\kappa_+}\left(\partial_t-\frac{\beta}{\sigma\kappa_+-\beta}\partial_{\pstar}\right)\,,\\ \bar{H}_{\pm1}&=\frac{ie^{\mp\kappa_+t}}{\sqrt{q(1+q)}} \left(\pm\frac{q(1+q)}{q'}\partial_r-\frac{1+2q}{2\sigma\kappa_+}\partial_t-\frac{\left(\sigma\kappa_+-\beta(1+2q)\right)}{2\sigma\kappa_+(\sigma\kappa_+-\beta)}\partial_\eta\right)\,,
	\end{split}
\end{equation}
and their $SL(2,\mathbb{R})\times U(1)$ Casimir which reproduces the Klein-Gordon equation on the $(t,r,\eta)$ section reads
\begin{equation}\label{eq: LT KG casimir}
	\begin{split}
		\mathcal{\hat{H}}^2&=-\Bar{H}_0^2+\frac{1}{2}\left(\Bar{H}_1\Bar{H}_{-1}+\Bar{H}_{-1}\Bar{H}_1\right)-\lambda\partial_\pstar^2
        \,.
	\end{split}
\end{equation}

\subsubsection{Einstein--Maxwell Lense--Thirring}

To illustrate the robustness of our procedure, we carry out the matching for slowing rotating charged black holes \cite{Gray:2021roq} where the metric functions take the form
\begin{equation}
  N=1,\quad f=1+p_i=1-\frac{16\pi M}{(d-2)\Omega_{d-2}r^{d-3}}+\frac{\mathfrak{e}^2}{r^{2(d-3)} }\,.
\end{equation}
Here $\mathfrak{e}$ is the electric charge parameter, and the corresponding electromagnetic potential and tensor are given by
\begin{equation}
    A=-\sqrt{\frac{d-2}{d-3}}\frac{\mathfrak{e}}{r^{d-3}}\left[dt -\sum_{i=1}^m(a_i\mu_i^2d\phi_i +\frac{a_i^2\mu_i^2p_i}{r^2f}dr) \right]\,,\quad F=dA\,.
\end{equation}
These metric functions and vector potential solve the Einstein--Maxwell equations derived from the Lagrangian
\begin{equation}
   {\cal L}=\frac{1}{16\pi}(R-F_{ab}F^{ab}).
\end{equation}

The key difference, compared to the pure Einstein case, is that in this case we have two, finite radius, distinct horizons in the spacetime. 
To see this note that by setting $\mathfrak{m}=8\pi M/((d-2)\Omega_{d-2})$, we can write 
\begin{equation}
    \Delta=r^{d-2}f=\frac{1}{r^{d-4}}(r^{d-3}-r_+^{d-3})(r^{d-3}-r_-^{d-3})\,,\quad r_\pm=(\mathfrak{m}\pm\sqrt{\mathfrak{m}^2-\mathfrak{e}^2})^{1/(d-3)}\,. 
\end{equation}
Note that we require $\mathfrak{m}>\mathfrak{e}$ in order that we have two distinct horizons, i.e. the system is not extremal. 
This requirement sets the scale of the mass and horizon radius so that $M\omega^{d-3}\ll1\iff\omega r_+\ll1$.

Moreover, we get again an exact matching of $\Delta=\Deff$ with the coefficients
\begin{equation}
    s=\frac{1}{(d-3)^2}\,,\quad B=r_+^{d-3}\,\quad C=r_-^{d-3}\,.
\end{equation}
Putting this together gives
\begin{equation}\label{eq: Deff Max}
    q=\frac{1}{r_+^{d-3}-r_-^{d-3}} (r^{d-3}-r_+^{d-3})\,, \quad \Deff=\frac{(r_+^{d-3}-r_-^{d-3})^2}{r^{d-4} }q(q+1)\,.
\end{equation}

It is now convenient to write $q$ and $\Deff$ in terms of the surface gravities
\begin{equation}
    \kappa_\pm=\frac{ f'(r_+)}{2}=\frac{(d-3)(r_\pm^{d-3}-r_\mp^{d-3}) }{2r_\pm^{d-2}}\,.
\end{equation}
Then we have
\begin{align}
    q&=\frac{(d-3)}{2\kappa_+r_+^{d-2}}(r^{d-3}-r_+^{d-3})
    =\frac{(d-3)}{2\kappa_+r_+}(\tilde{r}^{d-3}-1)\,,
    \\
    \Delta&=\Deff=\left(\frac{2\kappa_+r_+}{(d-3)}\right)^2r_+^{d-2}\left(\frac{r_+}{r}\right)^{d-4} q(q+1)=\left(\frac{2\kappa_+r_+}{(d-3)}\right)^2r_+^{d-2}\tilde{r}^{4-d} q(q+1)\,,
\end{align}
where again we have introduced the dimensionless variable $\tilde{r}=r/r_+$.
Notice, $q$ and $\Deff$ have exactly the same form as in the Einstein case (see e.g. \eqref{eq: Deff Ein}) except that $r_-$ is now nonzero. When $r_-=0$ the prefactor in these expressions, $2\kappa_+r_+/(d-3)=1$. Additionally, having a nonzero $r_-$ means that the pole structure is much more reminiscent of the Kerr and 5$d$ Myers--Perry cases since the $r=0$ behaviour is distinct from the inner horizon.

Next the function $p_i=a_i(f-1)/r^2$ becomes
\begin{align}
    p_i&=\frac{a_i}{r^2}(-1+r^{2-d}\Deff)=\frac{a_i}{r^2}\left(-1 + \left(\frac{2\kappa_+r_+}{(d-3)}\right)^2\left(\frac{r_+}{r}\right)^{2(d-3)} q(q+1) \right)\nonumber\\
    &=-\frac{a_i}{r_+^2}\left(\frac{r_+}{r}\right)^{2(d-2)}\left[\left(\frac{r}{r_+}\right)^{2(d-3)} -\left(\frac{2\kappa_+r_+}{(d-3)}\right)^2q(q+1)\right]\nonumber\\
    &=-\Omega^i_+\tilde{r}^{\,-2(d-2)}\left[ \tilde{r}^{2(d-3)} -\left(\frac{2\kappa_+r_+}{(d-3)}\right)^2q(q+1)\right]\,,
\end{align}
where as above $\Omega^i_+=-p_i(r_+)=a_i/r_+^2$.
Now, as for the pure Einstein case, let us write the non-derivative terms explicitly and isolate their pole structure. This gives,
\begin{align}\label{eq: Non deriv EMLT}
    \frac{sr^{d}}{\Delta}\bigl(\partial_t-\sum_i p_i\partial_{\phi_i}\bigr)^2
    =&\frac{sr^{d}}{\Deff}\bigl(\partial_t-\sum_i p^E_i\partial_{\phi_i}\bigr)^2
    \nonumber\\
    =&\frac{1}{(2\kappa_+)^2q(q+1)}\tilde{r}^{2(d-2)}\nonumber\\
    &\times
        \left\{ \partial_t +\tilde{r}^{-2(d-2)}\left[\tilde{r}^{2(d-3)}
         -\left(\frac{2\kappa_+r_+}{(d-3)}\right)^2q(q+1)\right]\sum_i\Omega^i_+\partial_{\phi_i}\right\}^2\nonumber\\
    =&\left(\frac{1}{q}-\frac{1}{q+1}\right)\nonumber\\
    &\times\left\{\frac{1}{2\kappa_+}\left[\tilde{r}^{d-2}(\partial_t+\tilde{r}^{\,-2}\partial_\pstar) -\tilde{r}^{-(d-2)}\left(\frac{2\kappa_+r_+}{(d-3)}\right)^2q(q+1)\partial_\pstar\right]\right\}^2\nonumber\\
    =&\left(\frac{1}{q}-\frac{1}{q+1}\right)\left[\frac{1}{2\kappa_+}
    \tilde{r}^{d-2}
    (\partial_t+\tilde{r}^{\,-2}\partial_\pstar)\right]^2
    \nonumber\\ 
   &-\frac{2r_+^2}{(d-3)^2}\, 
   (\partial_t+\tilde{r}^{\,-2}\partial_\pstar)\partial_\pstar
    \nonumber\\
    &+
   \tilde{r}^{-2(d-2)}\left(\frac{2\kappa_+r_+^2}{(d-3)^2}\right)^2
   q(q+1)\, 
   \partial_\pstar^2
\end{align}

Before proceeding we make the remark that there is a relation between the surface gravities and angular velocities at the inner and outer horizons. That is,
\begin{equation}\label{eq: pm horizon relations}
   \kappa_+r_+^{d-2}=-\kappa_-r_-^{d-2}\,,\quad r_+^2\Omega^i_+=r_-^2\Omega^i_-\,.
\end{equation}
 Then recalling from \eqref{killinggen} that $\partial_\pstar=\sum_i\Omega^i_+\partial_{\phi_i}$ and introducing $\partial_{\pstar_-}=\sum_i\Omega^i_-\partial_{\phi_i}$, where the combination $\partial_t+\partial_{\pstar_-}$ generates the inner horizon, we have a further symmetry between $q$ and $q+1$. 
 That is, we can write
\begin{equation}
    q+1=\frac{(d-3)}{-2\kappa_-r_-}(\check{r}^{d-3}-1)\,,\quad
    \tilde{r}^{-2}\partial_\pstar=\check{r}^{-2}\partial_{\pstar_-}\,
\end{equation}
where we have defined $\check{r}=r/r_-$. 
Thus we can analyse the $1/q$ terms and immediately obtain the expressions for the $1/(q+1)$ terms by sending $(\tilde{r},\partial_\pstar,\kappa_+)\to(\check{r},\partial_{\pstar_-},-\kappa_-)$.
We have,
\begin{align}\label{eq: q=0 pole EMLT}
    \frac{1}{q}\left[\frac{1}{2\kappa_+}
    \tilde{r}^{d-2}
    (\partial_t+\tilde{r}^{\,-2}\partial_\pstar)\right]^2
    =&\frac{1}{q}\left[\frac{1}{2\kappa_+}
    (\partial_t+\partial_\pstar)\right]^2
    +\frac{1}{(2\kappa_+)^2}\left(\frac{2\kappa_+r_+}{(d-3)}\right)
    \nonumber\\
    &\times
    \left( 
    \frac{\tilde{r}^{2(d-2)}-1}{\tilde{r}^{d-3}-1} \partial_t^2
    +\frac{\tilde{r}^{2(d-2)-2}-1}{\tilde{r}^{d-3}-1} 2\partial_t\partial_\pstar
    +\frac{\tilde{r}^{2(d-2)-4}-1}{\tilde{r}^{d-3}-1} \partial_t^2
    \right)
    \nonumber\\
    =&\frac{1}{q}\left[\frac{1}{2\kappa_+}
    (\partial_t+\partial_\pstar)\right]^2 
    \nonumber\\
    &+\frac{r_+}{2\kappa_+(d-3)}
    \left( 
    \frac{1+\sum_{k=1}^{2d-5}\tilde{r}^k}{1+\sum_{k=1}^{d-4} \tilde{r}^k} \partial_t^2
    \right.
    \nonumber\\
    &\left.
    \quad +\frac{ 1+\sum_{k=1}^{2d-7}\tilde{r}^k}{1+\sum_{k=1}^{d-4}\tilde{r}^k } 2\partial_t\partial_\pstar
    +\frac{1+\sum_{k=1}^{2d-9}\tilde{r}^k}{1+\sum_{k=1}^{d-4}\tilde{r}^k } \partial_\pstar^2
    \right)\,.
\end{align}
And thence
\begin{align}\label{eq: q=1 pole EMLT}
    -\frac{1}{q+1}\left[\frac{1}{2\kappa_+}
    \tilde{r}^{d-2}
    (\partial_t+\tilde{r}^{\,-2}\partial_\pstar)\right]^2
    =&-\frac{1}{q+1}\left[\frac{1}{2\kappa_-}
    \check{r}^{d-2}
    (\partial_t+\check{r}^{\,-2}\partial_{\pstar_-})\right]^2
    \nonumber\\
    =&-\frac{1}{q+1}\left[\frac{1}{2\kappa_-}
    (\partial_t+\partial_{\pstar_-})\right]^2 
    \nonumber\\
    &
    +\frac{r_-}{2\kappa_-(d-3)}\left( 
    \frac{1+\sum_{k=1}^{2d-5}\check{r}^k}{1+\sum_{k=1}^{d-4} \check{r}^k} \partial_t^2
    \right.
    \nonumber\\
    &\left.
    \frac{ 1+\sum_{k=1}^{2d-7}\check{r}^k}{1+\sum_{k=1}^{d-4}\check{r}^k } 2\partial_t\partial_{\pstar_-}
    +\frac{1+\sum_{k=1}^{2d-9}\check{r}^k}{1+\sum_{k=1}^{d-4}\check{r}^k } \partial_{\pstar_-}^2
    \right)\,.
\end{align}
Therefore, we have identified the residues of the $r_\pm$ poles. 
Keeping with our philosophy we will exactly match the $r_+$ pole and drop terms which are small in the near-region $\omega r_+\ll1$ and small near $r=r_+$. 
Substituting \eqref{eq: q=0 pole EMLT} and \eqref{eq: q=1 pole EMLT} back into \eqref{eq: Non deriv EMLT} and then using \eqref{eq: pm horizon relations} we have
\begin{align}
    \frac{sr^{d}}{\Delta}\bigl(\partial_t-\sum_i p_i\partial_{\phi_i}\bigr)^2
     =&\frac{1}{q}\left[\frac{1}{2\kappa_+}\left(\partial_t+\partial_\pstar\right)\right]^2 -\frac{1}{q+1}\left[\frac{1}{2\kappa_-}\left(\partial_t+\partial_{\pstar_-} \right)\right]^2
     \nonumber\\
     &+\left(\frac{(d-4)r_+}{(d-3)^2\kappa_+}
     -\frac{1}{4\kappa_+^2}\left(1-\left(\frac{r_-}{r_+}\right)^{2(d-4)}\right)
     -\frac{2r_+^2}{(d-3)^2}
     \right)\partial_\pstar^2
     \nonumber\\
     &+O(\omega r_+) +  O(r-r_+)\,,
\end{align}

Now, we match the $r=r_+$ ($q=0$) pole exactly by solving the equation
\begin{align}
    \left[\frac{1}{2\kappa_+}\left(\partial_t+\partial_\pstar\right)\right]^2+O(\omega r_+) +  O(r-r_+)&=\left(\frac{(\gamma+\alpha)\partial_t-(\beta+\delta)\partial_{\eta}}{2(\beta\gamma-\alpha\delta)}\right)^2
\end{align}
This is exactly the same equation as for the Kerr case \eqref{KerrRPpole} (with the appropriate modification of the surface gravity). 
Thus, setting $\gamma=0$ and matching the $r_+$, i.e. $q=0$, pole to leading order,  we are again forced to identify (up to choice of sign $\sigma=\pm1$)
\begin{equation}\label{eq: RNLT params}
\delta=-\sigma\kappa_+, \qquad \alpha=\sigma\kappa_+-\beta\,.
\end{equation}

Moreover to match as closely as possible the remaining $q=-1$ ($r=r_-$) pole terms, after one has substituted the solution of the parameters \eqref{eq: RNLT params}, it remains to identify $\lambda$ from the following equation:
\begin{align}
   \lambda\partial_\eta^2 =& -\frac{(\partial_t (\beta -\kappa_+ \sigma )+\partial_\eta (\beta +\kappa_+ \sigma ))^2}{4 \kappa_+^2 \sigma ^2 (\beta -\kappa_+ \sigma )^2}
    +\left[\frac{1}{2\kappa_-}\left(\partial_t+\partial_{\pstar_-}\right)\right]^2
    \nonumber\\
     &-\left(\frac{(d-4)r_+}{(d-3)^2\kappa_+}
     -\frac{1}{4\kappa_+^2}\left(1-\left(\frac{r_-}{r_+}\right)^{2(d-4)}\right)
     -\frac{2r_+^2}{(d-3)^2}
     \right)\partial_\pstar^2
     +O(\omega r_+) +  O(r-r_+)\,.
\end{align}
Thus we find
\begin{equation}
    \lambda=-\frac{\sigma\beta}{\kappa_+(\beta-\sigma\kappa_+)^2}-\left(
     \frac{(d-4)r_+}{(d-3)^2\kappa_+}
    -\frac{2r_+^2}{(d-3)^2}
    \right)\,.
\end{equation}
Notice that in $d=4$ this reduces to the same as the 4d Kerr case \eqref{eq: lambda}. Moreover, if we send $\mathfrak{e}\to0\iff r_-\to0$ then we recover the Einstein Lense--Thirring result \eqref{eq: lambda LT}.
To conclude, for Einstein--Maxwell Lense--Thirring black holes the generators are the same as for Einstein gravity \eqref{eq: LT generators}, up to the appropriate modification of ther surface gravity and $\lambda$, and the Casimir matching to the Klein--Gordon equation is again \eqref{eq: LT KG casimir}.

These two examples share a similar structure modulo, how in the Einstein case without Maxwell, the $r=0$ pole had to be treated carefully. However, in other theories of gravity, one would not expect to have an exact matching of the radial derivatives. 
Further, the extra parameters (of the particular theory one considers) could lead to the overall factor $s$ being modified in the near-region to leading order in these additional parameters. 
We leave this to future work.

Finally in the next section we show that these Lense--Thirring black holes allow for a separation of variables in infalling coordinates which may provide an alternative framework to study near horizon physics.

\subsection{Infalling coordinates}\label{sec: infalling}
The metric is regular on the horizon, and near its vicinity admits the Painlev{\'e}--Gullstrand (PG) {form
\cite{Baines:2021qaw,Gray:2021toe,Gray:2021roq}.
} 
Under the following coordinate transformation:
\ba\label{eq: PG coords}
dt&=&dT-\sqrt{\frac{1-f}{N}}\frac{dr}{f}\,, \nonumber\\
d\phi_i&=& d\Phi_i+p_i
\sqrt{\frac{1-f}{N}}\frac{dr}{f}\,,
\ea
 we recover
\be\label{LTHD2}
ds^2=-NdT^2+\Bigl(dr+\sqrt{N(1-f)}dT\Bigr)^2+r^2\sum_{i=1}^m \mu_i^2\Bigl(d\Phi_i+p_i dT\Bigr)^2
+r^2(\sum_{i=1}^{m+\epsilon}\!d\mu_i^2)\,,
\ee
which is manifestly regular on the horizon, 
and  the $T=const$. slices are manifestly flat. 
The inverse of the metric in the PG coordinates can be written
\begin{equation}\label{LTH inv D2}
    \partial_s^2= -\frac{1}{N}\left(\partial_T-\sqrt{N(1-f)}\partial_r -\sum_{i=1}^{m}p_i\partial_{\phi_i} \right)^2 +\partial_r^2 +\frac{1}{r^2}\left(\sum_{i=1}^m\frac{1}{\mu_i^2} \partial_{\phi_i}^2 +\sum_{\mu=1}^{m+\epsilon-1}\frac{U_\mu}{X_\mu}\partial_{x_\mu}^2\right).
\end{equation}

We can exploit the inverse metric and the fact that the coordinate transformation \eqref{eq: PG coords} does not change the determinant to separate the scalar wave equation in these coordinates. 
Repeating the steps in section \ref{subsec: LT Sep} with the ansatz
\be
\Psi=e^{-i\tilde{\omega} T+i\sum_{i=1}^{m}\tilde{m}_i\Phi_i} R(r)\prod_{\mu=1}^{m+\epsilon-1} S_\mu(x_\mu)\,,
\ee
we obtain the following separated equations.
Only the radial equation changes and it is given by,
\begin{multline}\label{eq: LT rad PG}
   \frac{r^2}{g_rR}\left[g_r\left\{f R'-i\left(\sqrt{\frac{1-f}{N}}[\tilde{\omega} +\sum_i \tilde{m}_ip_i(r)]R  \right)\right\}\right]'\\
   -ir^2\sqrt{\frac{1-f}{N}}[\tilde{\omega} +\sum_i \tilde{m}_ip_i(r)]\frac{R'}{R}+\frac{r^2}{N}\left(\sum_i \tilde{m}_i p_i(r) +\omega \right)^2-K=0\,, 
\end{multline}
The angular ones are the same up to shifting $m\to\tilde{m}$ are
\be
\frac{1}{S_\mu g_\mu}\Bigl({X_\mu g_\mu}\partial_{x_\mu}S_\mu\Bigr)_{\mu}-\frac{\tilde{m}_\mu^2}{x_\mu^2} +K_{\mu-1}= \frac{ K_\mu }{X_\mu}\,.
\ee
Thus the separation constant remains $K=\ell(\ell+d-3)$.

This radial equation may be useful for studying near horizon behaviour of solutions due to the regularity of the PG coordinates. 
Note, such an equation does not exist for Kerr black holes since there we do not have PG coordinates~\cite{Visser:2022fwx}. 
Moreover, it is the special nature of these Lense--Thirring spacetimes where $p_i$ is only a (regular) function of $r$, which means that \eqref{eq: LT rad PG} only depends on $r$.

\section{Discussion} \label{sec: diss}

In this work we developed a systematic procedure for calculating globally-defined Love generators for a large class of black holes in four and higher dimensions. 
Namely, we considered black hole backgrounds that admit a separable Klein--Gordon equation, and whose outer horizon is a Killing horizon generated by the Killing vector field \eqref{xi}.
We began by demonstrating that our procedure reproduces the known results for the Love generators of 4D Kerr and 5D Myers--Perry black holes, upon setting our conformal coordinate parameter $\beta=0$. 
We then moved on to our higher-dimensional black holes of interest, the Lense--Thirring black holes, which model higher-dimensional slowly rotating black holes. 
We showed for the first time that Lense--Thirring spacetimes allow a radial separation of variables, as in \eqref{KGIsolatingK}. 
We then built the Love generators for Einstein and Einstein--Maxwell Lense--Thirring spacetimes, and shown that they are strikingly similar in form to those of the 4D Kerr and 5D Myers--Perry black holes.

The Love generators are determined by our systematic procedure (Section \ref{sec: Conf Symmetry}) for matching the radial scalar Klein--Gordon operator, $\nabla^2$, of the separable spacetime in question to the quadratic Casimir, $\mathcal{H}^2$, of a global symmetry group containing $SL(2,\mathbb{R})$. 
There are several elements of this matching procedure to highlight: 
\begin{itemize}[leftmargin=*]

    \item While most hidden conformal symmetry authors concentrate on matching the radial Laplacian to the $SL(2,\mathbb{R})$ Casimir, we find it useful to match to the full global symmetry group of the system, $SL(2,\mathbb{R})\times\mathcal{M}$, where $\mathcal{M}$ consists of commuting $U(1)$s that represent periodic angular directions. 
    For example, in 4D Kerr and higher-dimensional Lense--Thirring spacetimes we find $\mathcal{M}\equiv U(1)$, and in 5D Myers--Perry $\mathcal{M}\equiv U(1)^2$. 
    {The number of $U(1)$ factors is related to the number of independent angular parts of the generators of the horizons $\partial_\pstar$, which for the Lense--Thirring spacetimes are proportional to one another, whereas for {the generically spinning} 5D Myers--Perry they are not.}
    Considering the full global group $SL(2,\mathbb{R})\times\mathcal{M}$ strengthens our argument for the existence of a 1-parameter family of generators.
    
    \item It is important to note that the matching of $\nabla^2$ to $\mathcal{H}^2$ is valid up to an overall constant $s$. We discuss the significance of this constant in extracting tidal Love numbers below. 
    
    \item We make use of the near-region, \eqref{eq: near-region}, limit of \cite{Castro:2010fd}: $\omega r \ll 1$ and $M\omega^{d-3} \ll 1$.
    
    \item For higher-dimensional black hole spacetimes with more than 2 poles, we suggest that a stronger limit is necessary, namely expanding the radial Laplacian in orders of $r-r_+$. 
    We discuss this further below. 
\end{itemize}

 The constant $s$ plays a crucial role in extracting tidal Love numbers. Imposing global well-definedness of one set of the symmetry generators leads to the arguments in~\cite{Charalambous:2021kcz, Charalambous:2023jgq} which show that the Love numbers (describing the tidal deformation) of Kerr holes vanish but those of the 5$D$ Myers--Perry black do not unless the corresponding $\ell/2$ is an integer. 
 For higher-dimensional Lense--Thirring spacetimes, one can use the Casimir \eqref{eq: LT KG casimir} to solve the Casimir eigenvalue equation equation representing the effective Klein--Gordon system built from these generators (cf. \eqref{eq: eff KG}) to find the static Love numbers
\begin{equation}\label{eq: LT KG Cas match}
    {\cal \hat{H}}\Psi=s\ell(\ell+d-3)\Psi=\hat{\ell}(\hat{\ell}-1)\Psi\,.
\end{equation}
The ``effective'' eigenvalues $\hat{\ell}=\frac{\ell}{d-3}$  are no longer all integers because of the overall factor $s=1/(d-3)^2$ in the matching process. 
This suggest that the arguments of \cite{Hui:2020xxx,Kol:2011vg,Charalambous:2022rre,Charalambous:2023jgq} continue to imply that the higher-dimensional Love numbers are generically nonzero. 
As the eigenvalues of the Klein--Gordon equation for the Lense--Thirring metrics are independent of the metric functions this conclusion seems robust to theories beyond Einstein gravity, although the particular details of the generators will change. 

In all of the spacetimes that we consider, the metric function responsible for the horizons, $\Delta$, has had a structure which allows for an exact matching of the derivative terms, i.e. $\Delta=\Deff$, c.f. \eqref{eq: Deff}. 
For slowly rotating black holes in generic theories this exact matching will no longer hold.
To apply our methods should however be straightforward (if messy in details) because the hidden conformal symmetry emerges in the near-region limit. 
Thus one would be able to match $\Deff$ to second order in a near horizon expansion. 
A further complication presents itself when the factor in the matching of the Casimir to the Klein--Gordon equation \eqref{eq: eff KG} is no longer constant, $s=s(r)$, such as in the Kerr-NUT A(dS) family for higher dimensions.
In this case, we could try expanding $s=s_0+s_1(r-r_+)+\dots$, checking to see if such a procedure provides physically meaningful matching.

In comparing the conformal generator parameters in 4D Kerr \eqref{kerrconditions}, 5D Myers--Perry \eqref{mpconditions}, general dimension Einstein Lense--Thirring \eqref{einltconditions}, and general dimension Einstein--Maxwell Lense--Thirring \eqref{eq: RNLT params} spacetimes, we see that the structure of Love generators is rather universal. 
Indeed, it is intriguing to use this procedure to study Love symmetry for other types of horizons, such as cosmological horizons. 
We leave this for future work. 

The methods introduced here, namely the generic conformal generators \eqref{moregengenerators}, the requirement for a resulting \emph{global} symmetry, and the stipulation to exactly match the outer horizon pole of the radially separated Klein--Gordon equation, are robust enough to be applied to the general-dimensional Kerr--NUT--(A)dS class of black holes~\cite{Gibbons:2004js} which have the required separability structure~\cite{Frolov:2017kze} for \eqref{KGIsolatingK}. 
However, the matching procedure has the same difficulties as mentioned in the previous paragraph. 
In particular, the radial equation involves a sum over all of the separability constants, corresponding to the extra Killing tensors present in higher dimensions, with non trivial $r$ dependence. 
This dependence would result in a non constant $s$ and so, as discussed above, the near horizon expansion is required even to match radial derivative terms.
We leave this to future study.

In our work we have also shown for the first time that the scalar field equation in generalized Lense--Thirring spacetimes admits separation of variables in Painlevé--Gullstrand coordinates. 
Such coordinates are associated with a freely infalling observer who registers flat space around her all the way to the singularity. 
These coordinates open an interesting opportunity for studying the near horizon hidden symmetries far away from the bifurcation surface, in coordinates that are regular along the entire black hole horizon. Importantly, coordinates with this regularity are not known for fully spinning black holes such as Kerr or Myers--Perry, see \cite{Visser:2022fwx}. 
This technical result may thus cast new light on the emergence of hidden symmetries in the presence of slow rotation and the physics of the near-region limit.

Although we only consider separable metrics in this work, it would be very interesting to apply our procedure to non-separable metrics. 
For example, hidden conformal symmetry was recently studied for 5D black ring solutions in \cite{Chanson:2022wls}. 
Even though these spacetimes are famously non-separable, in the near-region limit of \cite{Castro:2010fd} they are. 
Studying hidden conformal symmetry of more generic non-separable systems could shed light on a potential relationship between the hidden symmetries of separability (whose existence depends upon a tower of Killing--Yano tensors) and the hidden \emph{conformal} symmetries in the effective Laplacian. We leave this for future work.

\acknowledgments

The authors thank Francesco Di Filippo, Luca Iliesu, Sameer Murthy, Rahul Poddar, Mukund Rangamani, Maria Rodriguez, Adam Solomon, and Joaquin Turiaci for helpful discussions.  CK is supported by the U.S. Department of Energy under grant number DE-SC0019470 and by the Heising-Simons Foundation ``Observational Signatures of Quantum Gravity'' collaboration grant 2021-2818. CK also thanks the Aspen Center for Physics, which is supported by National Science Foundation grant PHY-2210452, for hospitality during the later stages of this project.
The authors thank the Centro de Ciencias de Benasque for hosting us during a portion of this project.
D.K. is grateful to GA{\v C}R 23-07457S grant of the Czech Science Foundation and  the Charles University Research Center Grant No. UNCE24/SCI/016 for their support.
F.G. acknowledges the Institute of Theoretical Physics at Charles University, and in particular, would like to thank D.K., Pavel Krtouš, and Lenka Knotková for their hospitality and organization hosting him while various aspects of this work were undertaken. 

\appendix

\section{Further properties of Lense--Thirring spacetimes}\label{Appendix: LT}

\subsection{The Killing tower}
In addition  to the $1+(m+\epsilon)$ Killing vectors $\partial_t$ and $\partial_{\phi_j}$ mentioned in the main text, the metric \eqref{LTHDimprovedApp} also admits a rapidly growing tower of Killing tensors.
Following \cite{Gray:2021toe,Gray:2021roq}, let us review here the corresponding construction. 
It starts from defining the following auxiliary objects:%
\footnote{``Mimicking'' the construction of hidden symmetries in Kerr-NUT-AdS spacetimes, e.g. \cite{Frolov:2017kze}
}
\begin{align}
	2b^{(I)}&\equiv r^2(dt+\sum_{i\in I}a_i\mu_i^2d\phi_i)\,,\quad\,h^{(I)}\equiv db^{(I)}\,,\\
	f^{(I)}&\equiv \frac{\sqrt{N}}{(|I|+1)!}*\big(\underbrace{h^{(I)}\wedge \dots \wedge h^{(I)}}_{|I|+1\ \mbox{\tiny times}}\big)\,.
\end{align}
Here, the index $I$ belongs to the power set of the set $S=\{1,..,m\}$, $I\in P(S)$, i.e. the set of all subsets of $S$, and $|I|$ denotes the corresponding size of $I$. 
One should think of this as a choice of rotation directions for which one wishes to construct a symmetry.

It turns out that the above auxiliary objects then define exact rank-2 Killing  tensors:
\begin{align}
K^{(I)}_{ab}&=\big(\prod_{i\in I
.
} a_i\big)^{-2}\,(f^{(I)}\cdot f^{(I)})_{ab}\,,\quad 
 K^{(I)}_{(ab;c)}=0\,,
\end{align}
where $ 
(\omega \cdot \omega)_{ab}=\frac{1}{p!}\omega_{ac_1\dots c_p}\omega_b{}^{c_1\dots c_p}
$
for any $(p+1)$-form $\omega$. Using the explicit form of the Lense--Thirring spacetimes \eqref{LTHDimprovedApp}, these Killing tensors can be written as follows:
\begin{equation}\label{KTs}
K^{(I)}=\sum\limits_{i\not\in I}^{m-1+\epsilon}\bigg[\bigr(1-\mu_i^2-\!\sum_{j\in I}\mu_j^2\bigr)(\partial_{\mu_i})^2 -2\!\!\sum_{j\not\in I\cup\{i\}}\!\! \mu_i\mu_j\,\partial_{\mu_i}\partial_{\mu_j} \bigg] 
+\sum\limits_{i\not\in I}^{m}\bigg[\frac{1-\sum_{j\in I}\mu^2_j}{\mu_i^2}(\partial_{\phi_i})^2\bigg] \,.
\end{equation}
While many of these are reducible, one can find a quadratically growing (with the number of dimensions) subset of $\frac{1}{2}(m-1+2\epsilon)$ Killing tensors that are irreducible (cannot be reduced to a sum of products of Killing vectors), provided $p_i=p_i(r)$ are sufficiently generic. 
A subset of 
\be 
\tilde m\equiv (m-1+\epsilon) 
\ee
of these can be shown to mutually commute and, together with the explicit Killing symmetries $\partial_t$ and $\partial_{\phi_i}$, they guarantee separability of the Hamilton--Jacobi \cite{Sadeghian:2022ihp} and (as explicitly demonstrated in this paper) also the Klein--Gordon equation, see next section as to why this is possible and \cite{Gray:2021toe,Gray:2021roq} for more details on these remarkable towers of hidden symmetries.  

\subsection{Standard separable form of Lense--Thirring metrics}\label{Appendix: LT 2}

{In this appendix we show that in the newly introduced coordinates which demonstrate that the angular part of the Lense--Thirring metrics the line element \eqref{metric} is actually in the standard separable form of Benenti~\cite{Benenti:1979} and moreover satisfies the Robinson condition~\cite{Robertson:1928} which guarantees the separability of the Klein--Gordon equation.

To do this, we group the coordinates  into $y^a=(y^A,y^\alpha)$ where the metric only depends on the coordinates $y^A=(r,x_\mu)$ for $A=0,\dots m+\epsilon-1$, and the Killing coordinates are denoted $y^\alpha=(t,\phi_i)$ for $\alpha=0,\dots,m+1$. 

If we in addition define
\begin{equation}\label{eq:Ueq2}
	Y_{A}=(f(r),X_\mu)\,,\quad V_{A}=(r^2,U_\mu)\,,
\end{equation}
then the inverse metric and one collection of $\tilde{m}=m+\epsilon-1$ mutually SN commuting Killing tensors corresponding to the nested sets 
\begin{equation}
\emptyset \subset\{1\} \subset\{1,2\} \subset\dots \subset\{1,\dots, \tilde{m}\}\,
\end{equation}
can be written in standard form:
\begin{equation}
	K_{(A)}=\sum_{B=0}^{m+\epsilon-1}\tensor{[\phi^{-1}]}{_A^B}\left(\partial_B\otimes\partial_B+[\zeta_{(B)}]^{\alpha\beta} \partial_\alpha\otimes\partial_\beta\right)\,,
\end{equation}
where we have introduced the matrix
\begin{equation}
\tensor{[\phi^{-1}]}{_A^B}=\sum_{C=0}^{m+\epsilon-1} \frac{V_AY_{B}}{V_{B}}  \, \tensor{\delta}{_{A}^{B-C}}\,.
\end{equation}
The first row of $\tensor{[\phi^{-1}]}{_A^B}$ corresponding to $A=0$ is the metric. 
This matrix $\tensor{[\phi^{-1}]}{_A^B}$ is also the inverse of the following St\"ackel matrix:
\begin{equation}
	\tensor{[\phi]}{_A^B}=\frac{V_A}{Y_AV_B}\left[\tensor{\delta}{_A^{B}}-\tensor{\delta}{_A^{B-1}}\right]=\frac{1}{Y_A}\left[\tensor{\delta}{_A^{B}}-\frac{1}{X_{A}}\tensor{\delta}{_A^{B-1}}\right]\,,
\end{equation}
where we set $X_{0}=r^2$. 
Finally, $[\zeta_{(B)}]^{\alpha\beta}$ is given by,
\begin{align}
	[\zeta_{(0)}]^{\alpha\beta}\partial_{\alpha}\partial_{\beta}&=-\frac{1}{Nf^2}\bigl(\partial_t-\sum_ip_i\partial_{\phi_i}\bigr)^2\,,\\
	[\zeta_{(\mu)}]^{\alpha\beta}\partial_{\alpha}\partial_{\beta}&=\frac{1}{x_\mu^2 X_\mu}(\partial_{\phi_\mu})^2\,\quad \text{for}\  \mu <\tilde{m}\,,\\
	[\zeta_{(\tilde{m})}]^{\alpha\beta}\partial_{\alpha}\partial_{\beta}&=\frac{1}{ x_{\tilde{m}}^2  X_{ \tilde{m} } } (\partial_{\phi_{\tilde{m}}})^2+(1-\epsilon)\frac{1}{X_{\tilde{m}}}(\partial_{\phi_m})^2\,.
\end{align}

From the definition of $Y_A$ and $X_A$ \eqref{eq:Ueq2} it follows that the $A$-th row of the matrix $\tensor{[\phi]}{_A^B}$ only depends on the particular coordinate $y^A$, i.e. where the index is fixed,  $\tensor{[\phi]}{_A^B}=\tensor{[\phi]}{_A^B}(y^{A})$. 
This is the defining feature of a St\"ackel matrix.
Likewise $[\zeta_{(A)}]^{\alpha\beta}$ depends only on the particular coordinate $y^A$.

This is the standard separable form of Benenti~\cite{Benenti:1979} which guarantees the separation of the Hamilton--Jacobi equation. 
The Klein--Gordon equation will be separable if in addition the Robinson condition~\cite{Robertson:1928} is satisfied. 
In these coordinates it reads~\cite{benenti:2002a,benenti:2002b},
\begin{align}\label{eq:Rob con}
    \partial_{a}\Gamma_{b}&=0\,, \quad\text{ if } a\neq b\, \quad \text{where }\Gamma_a=g_{a b}\,g^{c d}\tensor{\Gamma}{^b_c_d}\,.
\end{align}
Making use of the explicit form of the metric in standard coordinates we have $\Gamma_\alpha=0$ and
\begin{equation}
    \Gamma_A=-\partial_A\log g^{A A}+\frac{1}{2}\partial_A\log\det g\,, \quad \text{no sum on $A$.}
\end{equation}

Thus, the fact that the determinant~\eqref{eq: DetMet} and the inverse metric components $g^{AA}=Y_A$ are products of functions of one variable $y^A$ guarantees that \eqref{eq:Rob con} is satisfied. 
Hence the Klein--Gordon equation separates.}

This is of course consistent with the previous observation in~\cite{Gray:2021toe} that the above subset of mutually commuting Killing tensors also satisfies Carter's geometric criterion~\cite{Carter:1977pq} $\nabla_a(K_\gamma{}^{[\alpha}R^{\beta]\gamma})=0$.
Namely, that such tensors give rise to operators $\nabla^a K^{(I)}_{ab}\nabla^b$ which, together with the Killing vector operators ${\cal L}_j=\xi^{(j)}_a\nabla^a$, $\xi^{(j)}\in\{\partial_t,\partial_{\phi_j}\}$, yield a complete set of $d$ mutually commuting 
operators for the Klein--Gordon equation whose common eigenfunction (in appropriate coordinates) should be the separated solution, e.g.~\cite{Frolov:2017kze}.

\bibliography{workingbib}
\bibliographystyle{JHEP}

\end{document}